\documentclass[modern]{aastex62}

\received{**}
\revised{**}
\accepted{**}
\submitjournal{ApJ}

\shorttitle{Extremely overionized plasma of G359.1$-$0.5}

\usepackage{threeparttable}
\usepackage{multirow}
\usepackage{color}

\begin{document}

\title{
Uniform distribution of the extremely overionized plasma associated with the supernova remnant G359.1$-$0.5}

\correspondingauthor{Hiromasa Suzuki}
\email{suzuki@juno.phys.s.u-tokyo.ac.jp}

\author{Hiromasa Suzuki}
\affiliation{Department of Physics, The University of Tokyo,
7-3-1 Hongo, Bunkyo-ku, Tokyo 113-0033, Japan}

\author{Aya Bamba}
\affiliation{Department of Physics, The University of Tokyo,
7-3-1 Hongo, Bunkyo-ku, Tokyo 113-0033, Japan}
\affiliation{Research Center for the Early Universe, The University of Tokyo, 
7-3-1 Hongo, Bunkyo-ku, Tokyo 113-0033, Japan
}

\author{Rei Enokiya}
\affiliation{ Department of Physics, Nagoya University, 
Furo-cho, Chikusa-ku, Nagoya, Aichi 464-8601, Japan
}

\author{Hiroya Yamaguchi}
\affiliation{Institute of Space and Astronautical Science (ISAS),
Japan Aerospace Exploration Agency (JAXA),
3-1-1 Yoshinodai, Sagamihara, Kanagawa 252-5210, Japan}

\author{Paul P. Plucinsky}
\affiliation{Harvard-Smithsonian Center for Astrophysics,
60 Garden Street, Cambridge, MA 02138, USA
}

\author{Hirokazu Odaka}
\affiliation{Department of Physics, The University of Tokyo,
7-3-1 Hongo, Bunkyo-ku, Tokyo 113-0033, Japan}
\affiliation{Research Center for the Early Universe, The University of Tokyo, 
7-3-1 Hongo, Bunkyo-ku, Tokyo 113-0033, Japan
}



\begin{abstract}

We report on the results of our detailed analyses on the peculiar recombining plasma of the supernova remnant (SNR) G359.1$-$0.5, and the interacting CO clouds.
Combining {\it Chandra} and {\it Suzaku} data, we estimated the ionization state of the plasma with a careful treatment of the background spectrum. The average spectrum showed a remarkably large deviation of the electron temperature ($\sim$0.17 keV) from the initial temperature ($>$ 16 keV), indicating that the plasma is in a highly recombination-dominant state. On the other hand, the recombination timescale $({\it n_{\rm e} t})$ is comparable to those of the other recombining SNRs ($\sim4.2 \times 10^{11}$ cm$^{-3}$ s). We also searched for spatial variation of the plasma parameters, but found no significant differences.

Using $^{12}$CO($J$=2--1) data obtained with NANTEN2, we found a new, plausible candidate for the interacting CO cloud, which has a line-of-sight velocity of $\sim -$20 km s$^{-1}$.
This indicates that the SNR is located at a distance of $\sim$4 kpc, which is the foreground of the Galactic center, as previously reported.
The associated CO cloud does not show clear spatial coincidence with the nearby GeV/TeV emission, indicating that the origins of the GeV/TeV emission are likely unrelated to G359.1$-$0.5.

\end{abstract}



\section{Introduction} \label{intro}

Recent X-ray observations have found enhanced radiative recombination continua (RRCs) in the X-ray emitting plasmas in dozens of middle-aged to old supernova remnants (SNRs). Such RRCs indicate that the plasmas are presently recombining rather than ionizing, and called recombining plasmas (RPs; e.g.,\,\citealp{yamaguchi, ozawa, ohnishi}).
In such plasmas, the ionization temperature ($kT_{\rm z}$) is significantly higher than the electron temperature ($kT_\mathrm{e}$), implying that the plasma electrons have undergone a rapid cooling process.
Since the plasmas have not reached collisional ionization equilibrium (CIE), the recombination timescale ($n_\mathrm{e}t$: elapsed time of relaxation due to Coulomb collision after the rapid cooling, multiplied by an electron density) must be shorter than the characteristic timescale to reach CIE after the rapid cooling $\approx \sum^{Z} _{z=0} (S_{z} + \alpha_{z})^{-1}$ ($\sim 10^{13}$ cm$^{-3}$ s),
where $S_{z}$ and $\alpha_{z}$ are rate coefficients of ionization and recombination for an ion of the charge $z$ and the atomic number $Z$, respectively \citep{itomasai, smith10}.
The processes which generate RPs in SNRs are still unclear. However, interestingly, since all of the known recombining SNRs interact with molecular clouds (MCs), the origin of RPs is thought to be related to their circumstellar environments \citep{itomasai, kawasaki}).

As the origin of RPs, two principal scenarios have been proposed.
One is the rarefaction scenario \citep{itomasai}.
If a supernova explosion occurs in a dense circumstellar medium (CSM), the CSM is quickly shock-heated and ionized.
When the shock breaks out of the CSM and enters the low-density interstellar medium (ISM), $kT_{\rm e}$ is decreased rapidly by adiabatic cooling and an RP can be generated.

The other scenario to make an RP is thermal conduction with MCs \citep{kawasaki}. When a shock collides with cold MCs, the plasma electrons can be rapidly cooled down and the plasma can become recombining.
Considering these scenarios, investigating spatial distribution of the parameters of RPs is important to address the origin of RPs. There has been several works in which spatially-resolved X-ray analyses were conducted on recombining SNRs (e.g. IC443; W44: \citealp{matsumura18}; W49B: \citealp{zhou18}; \citealp{matsumura18}; \citealp{yamaguchi18}; W28: \citealp{okon18}).
In particluar, \cite{matsumura18} and \cite{yamaguchi18} discussed spatial correlation between $kT_{\rm e}$ and $n_{\rm e}t$ in order to identify the formation scenario of the RPs.

The SNR G359.1$-$0.5 is located toward the Galactic center (GC) region and contains an RP \citep{ohnishi}.
It has a radio shell of $\sim$12' in radius \citep{downes}.
The centrally-filled X-ray emission first discovered with {\it ROSAT} \citep{egger} indicates that this object is among the ``mixed-morphology'' class \citep{rho98}.
The ${\it ASCA}$ observation detected prominent emission lines at $\sim$1.9 keV and $\sim$2.6 keV, which were interpreted as Si-K (He-like) and S-K (H-like) lines, respectively \citep{bamba00}.
Hence, this SNR was recognized as a quite peculiar object, because the heavier elements tend to be less ionized compared to the lighter ones in a normal isothermal plasma, so that the measured spectra required a cooler plasma consisting only of silicon and a hotter one with over-abundant sulfur.

Observing with {\it Suzaku}, however, \cite{ohnishi} first discovered remarkably strong Si-K (H-like) RRC emission, which had been misinterpreted as S-K (H-like) line emission in the $ASCA$ study. They were able to explain the spectra phenomenologically with a one component plasma model with $kT_{\rm e}$ $\sim$0.24 keV and $kT_{\rm z}$ $\sim$0.77 keV. This may indicate that, in this RP, the deviation from CIE is much larger than any known RPs in SNRs.
However, the ionization state of the plasma is still unclear, because previous spectral analyses did not include nonequilibrium calculations and had large uncertainties in X-ray background estimation.
The spatial variations of the plasma parameters, which provide crucial information on the origin of the RP, are also unknown.

Previously this SNR was thought to be located in the GC region since \cite{uchida92a} suggested that CO clouds with high line-of-sight velocities of $-60$ to $-190$ km s$^{-1}$ are surrounding the SNR, and the H\,I absorption study of the radio shell showed that H\,I clouds also had high velocities of $-75$ to $-190$ km s$^{-1}$ \citep{uchida92b}.
However, there has been no evidence for their interaction with the SNR.
On the other hand, OH (1720 MHz) masers that were detected from several points in the radio shell show much lower velocities than those of the H\,I clouds around $-5$ km s$^{-1}$ \citep{yusef95}.
Since the detection of 1720 MHz OH masers in the vicinity of the SNR indicates evidence of the shock-cloud interaction, it is expected that any associated clouds will have a similar velocity of $\sim -$5 km s$^{-1}$, that is, the SNR is located in the foreground rather than the GC region.
Also, other emission lines from several molecules which seem to be interacting with the SNR, were revealed to have similar velocities to that of the OH masers \citep{lazendic02}.
Moreover, the Mouse, a pulsar which is in the vicinity of G359.1$-$0.5 and has a comparable absorption column density \citep{mori05}, was revealed to be located at the distance of 3 to 5 kpc by its dispersion measure \citep{camilo02}.
Therefore, G359.1$-$0.5 is likely to be located in the foreground of the GC region, although no interacting CO cloud has been confirmed.

Since the spatial distribution of the hadronic gamma-ray emission reflects the distribution of the nearby clouds, the information of GeV/TeV gamma-rays are also important to understand the environment of the SNR.
Although the GeV/TeV gamma-ray emission is also detected from the vicinity of this SNR (\citealp{aharo08}: \citealp{hui16}), it is unclear whether it is associated with the SNR.

In this paper, we report first spatially-resolved X-ray analysis of the G359.1$-$0.5 plasma using $\it Chandra$ and $\it Suzaku$ archival data in order to understand the processes which generate the RP, and an analysis on the interacting clouds using NANTEN and NANTEN2 data in order to confirm the distance to the SNR.
Observation details are summarized in section 2, and the results are presented in section 3.
We discuss the physical parameters of the RP, and the distance to the SNR in section 4.
Our conclusion is summarized in section 5.
Throughout this paper, errors in the text and tables represent a 90\% confidence level.

\section{Observation and Data Reduction}

\subsection{X-ray Observations with {\it Chandra} and {\it Suzaku}}

\begin{table}[t]
 \caption{Details of the observations with {\it Suzaku} and {\it Chandra}.}
  \centering
  \begin{threeparttable}
   \begin{tabular}{l  l  c   l   c} \hline \hline
     Satelite & Observation ID  & ({\it l, b})  & Start date & Effective exposure (ks) \\ \hline

      {\it Chandra} & 13807 & ($359^\circ_.0967, -0^\circ_.4888$) & 2012-11-01 & 89.4    \\ \hline      
      {\it Suzaku}  & 503012010 & ($359^\circ_.0947, -0^\circ_.4452$) & 2008-09-14 & 57.7    \\
                          & 502016010\tnote{*} & ($358^\circ_.9172, -0^\circ_.4785$) & 2008-03-02 & 70.5   \\
                    \hline
      
        \end{tabular}
        \begin{tablenotes}\footnotesize
	\item[*] Observation toward an adjacent region to the G359.1$-$0.5 used for background estimation.
	\end{tablenotes}
        
        \end{threeparttable}
   \label{obs_x}
\end{table}

We used archival data of {\it Chandra} \citep{weis96} and {\it Suzaku} \citep{mitsuda}.
Observation logs are shown in Table~\ref{obs_x}.

First, we summarize the procedure of the {\it Chandra} data reduction.
G359.1$-$0.5 was observed with the I0--I3 chips of the Advanced CCD Imaging Spectrometer (ACIS; \citealp{garmire97}). The data mode was {\it VFAINT}.
We processed the raw data following the standard data reduction methods ({\tt chandra\_repro} tool with an option {\it check\_vf\_pha=yes}), which include the correction for the charge transfer inefficiency.
We used CIAO (v4.10; \citealp{fruscione06}) and calibration database 4.7.8 for the data reduction.

Next, we summarize the details of the {\it Suzaku} data reduction.
G359.1$-$0.5 was observed with the X-ray Imaging Spectrometers (XISs; \citealp{koyama}) onboard {\it Suzaku}.
In order to extract the background spectrum from a region outside the radio shell as well, we also used an adjacent observation data (the last row in Table~\ref{obs_x}).

In both of the {\it Suzaku} observations, only three sets of onboard XIS0, 1, and 3 were operated. XIS1 is a back-illuminated (BI) CCD whereas the others are front-illuminated (FI) ones.
The XISs were operated in the normal-clocking full-window mode.
The spaced-row charge injection (SCI; \cite{uchiyama}) technique was performed for all of the XISs.
In the XIS data screening for both of the two observations, we eliminated the data acquired during the passage through the South Atlantic Anomaly (SAA), having elevation angles with respect to Earth's dark limb below 5$^{\circ}$, or with elevation angles to the bright limb below 20${^\circ}$ in order to avoid contamination by emission from the bright limb.
We reprocessed the data with the calibration database version 2016-04-01.
The redistribution matrix files and the ancillary response files for the XISs were generated with {\tt xisrmfgen}, {\tt xissimarfgen} \citep{ishisaki}, respectively.
In the spectral analysis, the XIS0 and XIS3 spectra were merged and averaged.

The data reduction tasks were executed with HEAsoft (v6.20; \citealp{heasarc14}).
In the spectral analysis, we used XSPEC (v12.9.1; \citealp{arnaud96}), and AtomDB 3.0.9 for both {\it Suzaku} and {\it Chandra} data.

\subsection{Molecular line observations with NANTEN and NANTEN2} \label{obs_radio}

We used the $^{12}$CO($J$=1--0) and $^{12}$CO($J$=2--1) datasets obtained with NANTEN and NANTEN2.
Observations of $^{12}$CO($J$=1--0) were carried out 1999 March to 2001 September by using the NANTEN telescope at Las Campanas in Chile with a position-switching mode \citep{miz04}. The half-power beam width (HPBW) at 115 GHz, the grid spacing, and the velocity resolution are 2\arcmin.6, 2\arcmin, and 1 km~s$^{-1}$, respectively. The final data smoothed with a Gaussian function achieved r.m.s. noise fluctuations of $\sim$0.19 K/ch in $T_\mathrm{mb}$ with 164\arcsec ~angular resolution. Observations of $^{12}$CO($J$=2--1) were carried out 2010 July to 2011 January by using the NANTEN2 telescope at Pampa La Bola, Atacama in Chile with an on-the-fly mode \citep{eno14}. The HPBW at 230 GHz and the velocity resolution are 90\arcsec, and 1 km~s$^{-1}$, respectively. The final data smoothed with a Gaussian function achieved r.m.s. noise fluctuations of $\sim$0.20 K/ch in $T_\mathrm{mb}$ with 108\arcsec ~angular resolution.
We used $^{12}$CO($J$=1--0) data only to derive masses and column densities of molecular clouds, because its angular resolution is coarser than that of the $^{12}$CO($J$=2--1) data.

\section{Results}

\subsection{X-ray images}

\begin{figure}[ht!]
\centering
\plotone{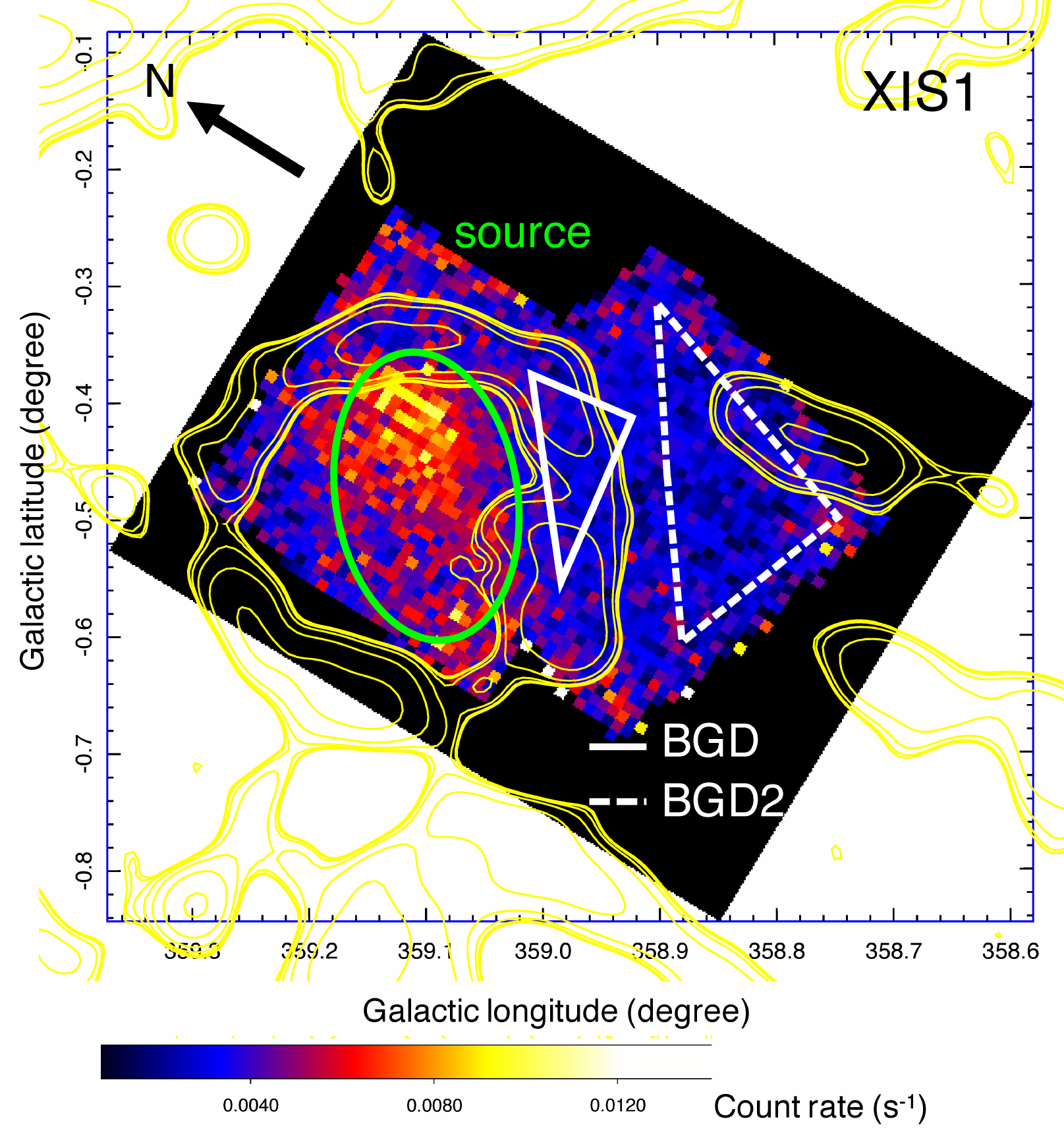}
\caption{ {\it Suzaku} X-ray image of G359.1$-$0.5 region (1.0--3.0 keV) shown in Galactic coordinate. Yellow contours represent 843 MHz flux observed with SUMSS. The green ellipse and white solid triangle represent the source and BGD regions for spectral analysis, respectively. The white dashed triangle represents the BGD2 region. The image is shown on a linear scale, and contours are on a logarithmic scale. The black arrow indicates the direction of north. \label{img_suzaku}}
\end{figure}

The 1.0--3.0 keV image extracted from the {\it Suzaku} data, in which the emission from G359.1$-$0.5 is dominant, is shown in Figure \ref{img_suzaku}.
The source and background (BGD) regions for our spectral analysis are shown with the green ellipse and white triangle. We also show the BGD2 region with a white dashed triangle outside the radio shell, which was used to evaluate the dependency of the plasma parameters on background assumption. The source region is roughly the same as the one used in \cite{ohnishi}.
The yellow contours represent SUMSS (Sydney University Molonglo Sky Survey; \citealp{bock99}) 843 MHz  flux \footnote{$\langle$https://skyview.gsfc.nasa.gov/current/cgi/titlepage.pl$\rangle$.}.
We can see the diffuse X-ray emission inside the radio shell.

\begin{figure}[ht!]
\centering
\plotone{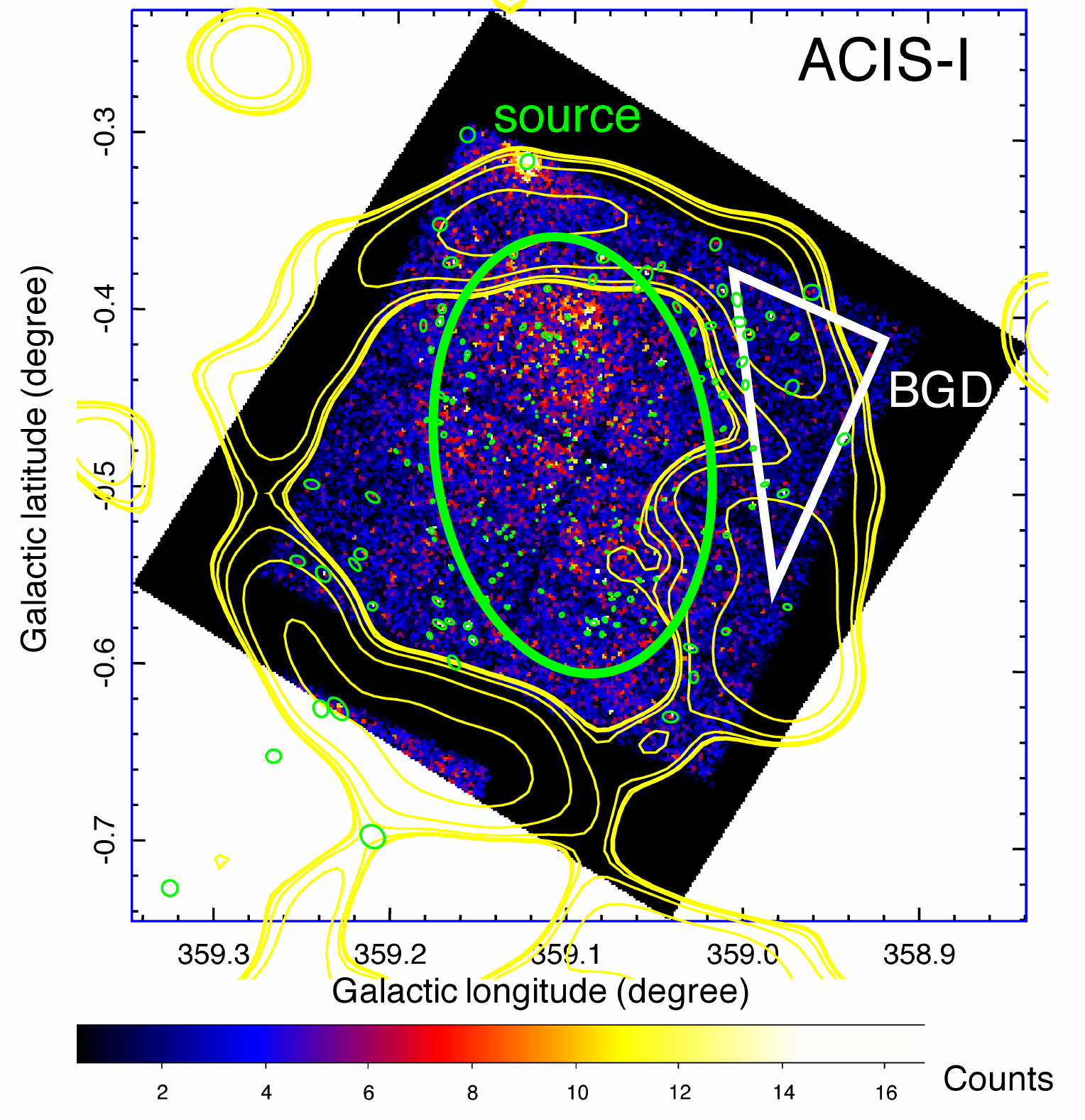}
\caption{ {\it Chandra} X-ray image of G359.1$-$0.5 region (1.0--3.0 keV). Yellow contours are the same as that in Figure \ref{img_suzaku}. The green, largest ellipse and white triangle represent the source and BGD regions for spectral analysis, respectively. The regions enclosed by green, tiny ellipses represent locations of point-like sources detected with the {\tt wavdetect} tool. The image is shown on a linear scale, and contours are on a logarithmic scale. \label{img_chandra}}
\end{figure}

Figure \ref{img_chandra} shows the X-ray image (1.0--3.0 keV) of G359.1$-$0.5 observed with {\it Chandra}. The source and BGD regions are the same as those in Figure \ref{img_suzaku}.
Point-like sources detected with {\it Chandra} analysis tool, {\tt wavdetect} are indicated with green, small ellipses.
We ran {\tt wavdetect} in the 0.7--5.0 keV energy range and with the parameters of $scales \,=\, 1, 2, 4, 8, 16$, and $sigthresh \,=\, 10^{-6}$.
The combined flux of the point source contribution was only $\sim$2\% and $\sim$1\% of the fluxes of the source and BGD regions, respectively. Thus we included the point sources in the spectral analysis for consistency with the $Suzaku$ spectra.
The bright source at the northwestern corner of the field of view is The Great Annihilator (1E1740.7-2942).

\begin{figure}[ht!]
\centering
\plotone{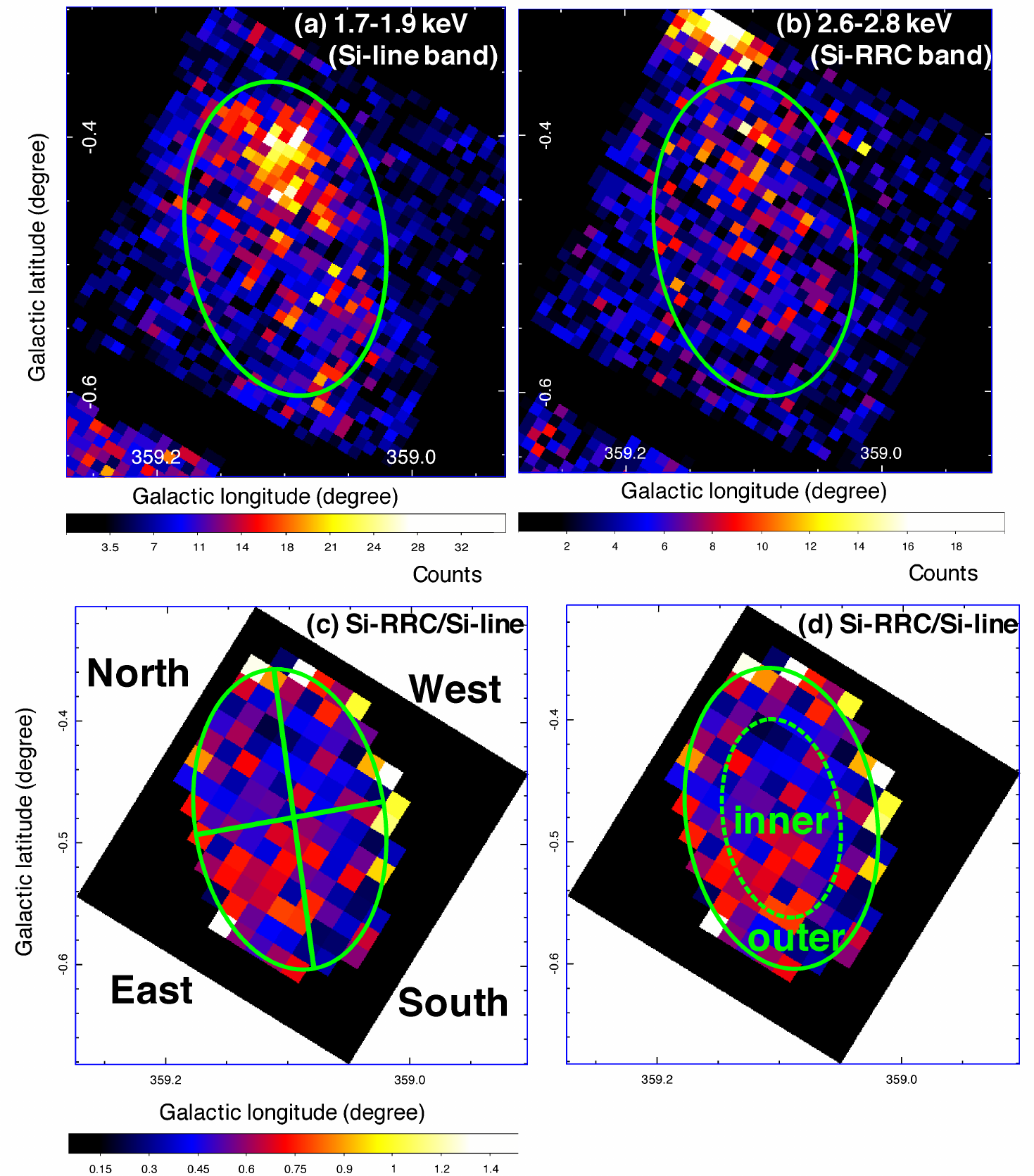}
\caption{ (a) The {\it Chandra} image of 1.7-1.9 keV energy band (Si-line band). (b) The {\it Chandra} image of 2.6-2.8 keV energy band (Si-RRC band).  (c) and (d) The image of flux ratio of Si-RRC band and Si-line band. The green solid ellipses in all the panels are the source region, which is the same as that in Figure \ref{img_chandra}. The four regions named North, West, East, and South regions are indicated in panel (c). In panel (d), the inner ellipse and outer oval ring are shown on the same image as panel (c). \label{img_chandra2}}
\end{figure}

In order to investigate spatial variations of the plasma parameters, we extracted images from the two energy ranges of the {\it Chandra} data, 1.7--1.9 keV (Si-line band) and 2.6--2.8 keV (Si-RRC band), which correspond to the Si-K line (He-like), Si-K RRC (H-like) emission, respectively.
Figure \ref{img_chandra2} (a) and (b) are the images of Si-line band and Si-RRC band, respectively. Figure \ref{img_chandra2} (c) shows the flux ratio of Si-RRC band and Si-line band. In these three images, the point-like sources detected above were excluded.
Figure \ref{img_chandra2} (a) and (b) show that the plasma is bright in the northwestern region, whereas Figure \ref{img_chandra2} (c) indicates that the RRC flux is relatively high in the eastern region.
We therefore divided the source region into four parts, North, South, West, and East as described in Figure \ref{img_chandra2} (c), to extract the spectrum from each region and compare the plasma parameters.
In order to check the parameter variations with radius, we also divided the source region into the inner ellipse and outer oval ring as shown in Figure \ref{img_chandra2} (d).

\subsection{Background estimation on {\it Chandra} and {\it Suzaku} data. }\label{bgd_spec}

In order to extract the emission from G359.1$-$0.5, we first estimated the background components.
The background has two components. One is X-ray background which originates from Galactic and extragalactic sources (Sky-BGD), and the other is detector background (Non X-ray Background: NXB).
We estimated these two components using the BGD-region data as follows.

First, we modeled the Sky-BGD component.
X-ray sources toward the GC region are significantly affected by Galactic background emission, especially by the Galactic ridge X-ray emission (GRXE). We applied the same modeling as that in \cite{nakashima13}, in which spectral analysis on regions near the BGD region in this paper were conducted: the model consists of a low-temperature GC plasma component (LP), a high-temperature GC plasma component (HP), and a foreground emission component (FE), all of which were modeled as CIE plasmas, and the neutral Fe-K line emission, which was modeled by a narrow Gaussian function with a fixed energy centroid of 6.4 keV.
The electron temperatures ($kT_{\rm LP}, kT_{\rm HP}, kT_{\rm FE}$) and the emission measures ($EM_{\rm LP}, EM_{\rm HP}, EM_{\rm FE}$) were treated as free parameters. The metal abundances were assumed to be the same for the LP and HP components. The abundances of Mg, Si, S, Ar, Ca, and Fe were treated as free parameters whereas the Ne and Ni abundances were tied to Mg and Fe, respectively. Abundances of the other elements were fixed to solar values. The abundances of Ne and Mg in the FE component were treated as free parameters while those of the other elements were fixed at solar values. The intensity of the Gaussian function was also included as a free parameter.
In addition to the Galactic background emission, we also considered the cosmic X-ray background (CXB) and NXB components. For {\it Suzaku} spectra, CXB was modeled by a power law (PL) model with a photon index of 1.4, and the flux was fixed at $6.38 \times 10^{-8}$ erg s$^{-1}$ cm$^{-2}$ str$^{-1}$ \citep{kushino02}.

The whole Sky-BGD model was therefore,
\begin{eqnarray}
&&{\rm Abs._{GC} \times (LP + HP + Neutral~Fe{\mathchar`-}K)} \nonumber \\
&&{\rm + Abs._{FE} \times FE + (2\,Abs._{GC}) \times CXB,}
\end{eqnarray}
where Abs.$_{\rm GC}$ (column density of $N_{\rm H, GC}$) and Abs.$_{\rm FE}$ (column density of $N_{\rm H, FE}$) represent interstellar absorptions toward the GC and FE components. The CXB component is affected by the absorption with the column density of twice that of $N_{\rm H, FE}$, since the CXB component propagates through the entire Galaxy.

Next, we describe the details of NXB estimation.
In the case of {\it Suzaku}, NXB spectra were estimated with the tool {\tt xisnxbgen} \citep{tawa08}, and subtracted from the spectral data.
On the other hand, in the spectroscopy using {\it Chandra} data, we determined the NXB model as follows.
The NXB spectrum includes a power-law (PL) component and several detector lines including Al-K$\alpha$ ($\approx$1.48 keV), Au-M$\alpha$ ($\approx$2.14 keV), Ni-K$\alpha$ ($\approx$7.48 keV), Ni-K$\beta$ ($\approx$8.27 keV), Au-L$\alpha$ ($\approx$9.71 keV) (CXC memo\footnote{$\langle$http://cxc.harvard.edu/cal/Acis/Papers/hist\_writeup.pdf$\rangle$}; \citealp{bartalucci14}).
We applied the NXB model using single PL and Gaussian components corresponding to these lines. The free parameters included the photon index and flux of the PL, and energy-centroids, line-widths and flux of the three of the Gaussian components (Au-M$\alpha$, Ni-K$\alpha$, Au-L$\alpha$). The energy-centroids and line-widths of Al-K$\alpha$ are fixed to 1.48 keV and zero, respectively and those of Ni-K$\beta$ are also fixed to 8.27 keV and zero, respectively, because of the inadequate statistics.

\begin{figure}[ht!]
\centering
\plotone{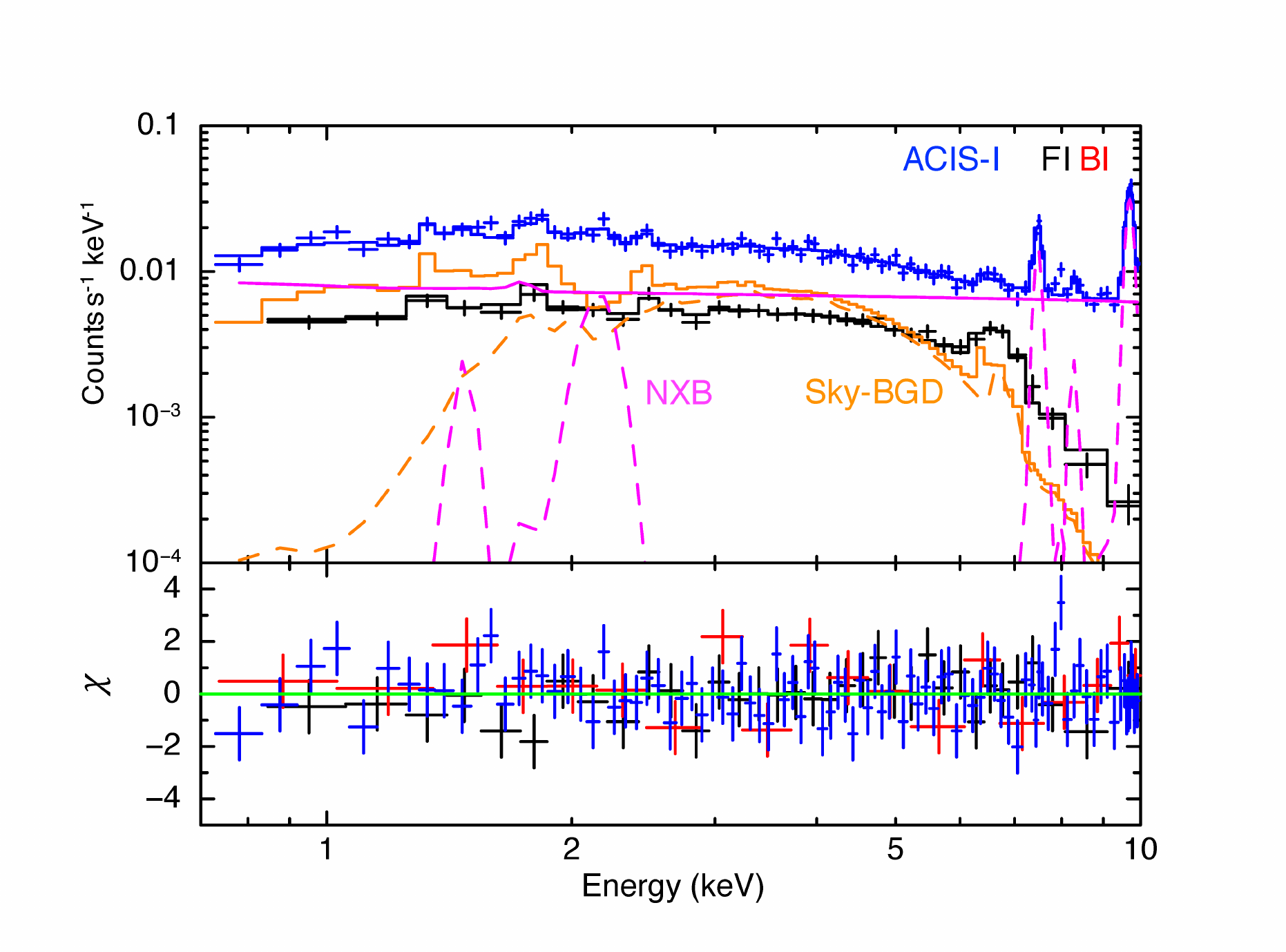}
\caption{The energy spectra, best-fit model and the residuals for BGD-region spectra obtained with {\it Chandra} and {\it Suzaku}. The blue crosses and line represent the {\it Chandra} ACIS-I data and the best-fit model, respectively. The black crosses and line represent those of {\it Suzaku} FI. For the {\it Suzaku} BI, only the residuals are shown with the red crosses. The orange, solid and dashed lines are the sum of the best-fit Sky-BGD model and the HP component for the {\it Chandra} ACIS-I, respectively. The magenta solid line and dashed lines are the PL and Gaussian components in the best-fit NXB model for the {\it Chandra} ACIS-I, respectively. \label{spe_bgd_sc}}
\end{figure}

Using Sky-BGD and NXB models described above, we fitted simultaneously the BGD-region spectra obtained with {\it Chandra} and {\it Suzaku}, in the 0.7--10.0 keV energy range.
We assumed that the Sky-BGD component in the BGD-region spectrum obtained with {\it Suzaku} is the same as that obtained with {\it Chandra}, assuming it to be time-invariant.
The BGD-region spectra, best-fit models, and the residuals are shown in Figure \ref{spe_bgd_sc}. The best-fit parameters are presented in Table \ref{tab_bgd_sc}.
The model fitted the data acceptably with reduced-$\chi^2 \,(d.o.f) = 1.04\,(123)$.
It is worth mentioning that we were able to get such a good fit with exactly the same Sky-BGD model on the {\it Chandra} and {\it Suzaku} data.

The best-fit parameters of the Sky-BGD component other than $N_{\rm H, FE}$ were not greatly different from those presented in \cite{nakashima13}.
As for $N_{\rm H, FE}$, the value obtained in this work ($\sim 0.95 \times 10^{22}~{\rm cm}^{-2}$) might be too large as the interstellar absorption for the foreground emission. If $N_{\rm H,FE}$ is fixed to zero or $N_{\rm H, FE}$ is fixed to the value in \citep{nakashima13} ($0.42 \times 10^{22}\, {\rm cm}^{-2}$), the model was also able to fit the data well with reduced-$\chi^2 \,(d.o.f) = 1.29\,(124)$ and 1.31\,(125), respectively. Also, the best-fit NXB parameters were not modified significantly from the ones in Table \ref{tab_bgd_sc}.
On the other hand, the best-fit parameters required over abundant Ne and Mg of the FE component ($>$1.1 and $>$2.4 (solar), respectively).
Thus, we were not able to determine which model was the most natural as the foreground emission.
However, it is important that such uncertainties do not affect our analysis since the NXB parameters are not affected by the FE parameters.

\begin{table*}[htb!]
\caption{ Best-fit parameters for BGD-region spectra. }
\centering
\begin{threeparttable}
\begin{tabular}{l l c} \hline\hline
Component & Parameter\tnote{*} & Value \\ \hline
Sky-BGD \\
FE& ${\it N}_{\rm H, FE}~{\rm (10^{22}~cm^{-2})}$ & 0.95 (0.88--1.12) \\
&${\it kT}_{\rm FE}~{\rm (keV)}$ & 0.13 (0.10--0.15) \\
&${\rm Ne~(solar)}$ & 0.05 ($<$ 0.25) \\
&${\rm Mg~(solar)}$ & 0 ($<$ 1.4) \\
&${\it EM}_{\rm FE}$ & 2.2 (1.0--27.4) \\
LP + HP + Neutral Fe-K& ${\it N}_{\rm H, GC}~{\rm (10^{22}~cm^{-2})}$ & 3.6 (2.9--4.3) \\
LP + HP& ${\rm Mg~(solar)}$ & 5.7 (1.6--24.4) \\
&${\rm Si~(solar)}$ & 1.4 (0.6--3.9) \\
&${\rm S~(solar)}$ & 3.2 (1.5--7.3) \\
&${\rm Ar~(solar)}$ & 4.0 ($<$ 11) \\
&${\rm Ca~(solar)}$ & 1.8 ($<$ 7.9) \\
&${\rm Fe~(solar)}$ & 0.9 (0.6--1.1) \\
LP& ${\it kT}_{\rm LP}~{\rm (keV)}$ & 0.48 (0.36--0.57) \\
&${\it EM}_{\rm LP}$ & 0.09 (0.03--0.32) \\
HP& ${\it kT}_{\rm HP}~{\rm (keV)}$ & 9.4 (7.8--11.6) \\
&${\it EM}_{\rm HP}$ & 0.036 (0.027--0.040) \\
Neutral Fe-K &  ${\it E}~{\rm (keV)}$ & 6.4 (fixed)\\
&$\sigma~{\rm (keV)}$ & 0 (fixed) \\
& flux (10$^{-4}$) & 1.45 (1.01--1.88) \\
\\
NXB\\
Al-K$\alpha$ & ${\it E}~{\rm (keV)}$ & 1.48 (fixed)\\
&$\sigma~{\rm (keV)}$ & 0 (fixed) \\
& flux & 0.0004 ($<$ 0.0007) \\
PL& ${\Gamma}$ & 0.10 (0.05--0.15) \\
&Norm (10$^{-3}$)\tnote{\dag} & 7.9 (7.2--8.6) \\
Au-M$\alpha$& ${\it E}~{\rm (keV)}$ & 2.16 (2.10--2.19) \\
&$\sigma~{\rm (keV)}$ & 0.09 (0.01--0.17) \\
&flux (10$^{-3}$) & 1.9 (1.2--2.7) \\
Ni-K$\alpha$& ${\it E}~{\rm (keV)}$ & 7.48 (7.46--7.49) \\
&$\sigma~{\rm (keV)}$ & 0 (fixed) \\
&${\it EM}_{\rm }$ (10$^{-3}$) & 3.0 (2.6--3.5) \\
Ni-K$\beta$ & ${\it E}$ (keV) & 8.27 (fixed) \\
&$\sigma~{\rm (keV)}$ & 0 (fixed) \\
& flux (10$^{-4}$) & 6.0 (2.9--9.1) \\
Au-L$\alpha$& ${\it E}~{\rm (keV)}$ & 9.69 (9.68--9.70) \\
&$\sigma~{\rm (keV)}$ & 0.054 (0.036--0.068) \\
&flux (10$^{-3}$) & 8.6 (7.9--9.3) \\
&{\it Reduced-}$\chi^2 / {\it d.o.f}$ & 1.04/123\\
\hline
\end{tabular}
\begin{tablenotes}\footnotesize
  \item[*] Emission measures ({\it EMs}) are shown in units of 10$^{-14}$(4$\pi${\it D}$^2$)$^{-1}$$\int${\it n}$_\mathrm{e}${\it n}$_\mathrm{H}${\it dV}~cm$^{-5}$, where {\it D}, {\it n}$_\mathrm{e}$ and {\it n}$_\mathrm{H}$ stand for distance (cm), electron and hydrogen number densities (cm$^{-3}$), respectively. Fluxes of Gaussian models are shown in units of cm$^{-2}$ s$^{-1}$.
\item[\dag] Normalization of the power-law model in units of cm$^{-2}$ s$^{-1}$ keV$^{-1}$ at 1 keV.

\end{tablenotes}
\end{threeparttable}
\label{tab_bgd_sc}
\end{table*}

\subsection{Spectroscopy on the entire source region with {\it Chandra} and {\it Suzaku}}\label{spe_source}

Using the BGD model obtained in section \ref{bgd_spec}, we conducted spectral fitting of the G359.1$-$0.5 plasma, in the 0.7--10.0 keV range.
Since it had been known to have an RP component, we used an absorbed RP model (${\tt vrnei}$ model in XSPEC) in addition to the BGD components to fit the source-region spectra.
The parameters in the RP model included an electron temperature $kT_{\rm e}$, initial temperature $kT_{\rm init}$, recombination timescale $n_{\rm e}t$, and metal abundances. This model assumed that a CIE plasma with the electron temperature of $kT_{\rm init}$ at the initial state subsequently decreased the electron temperature instantly to $kT_{\rm e}$, and became a recombination-dominant state (e.g. \citealp{foster14}; AtomDB 3.0 Documentation\footnote{$\langle$http://www.atomdb.org/atomdb\_300\_docs.pdf$\rangle$}).

\begin{figure}[ht!]
\centering
\plotone{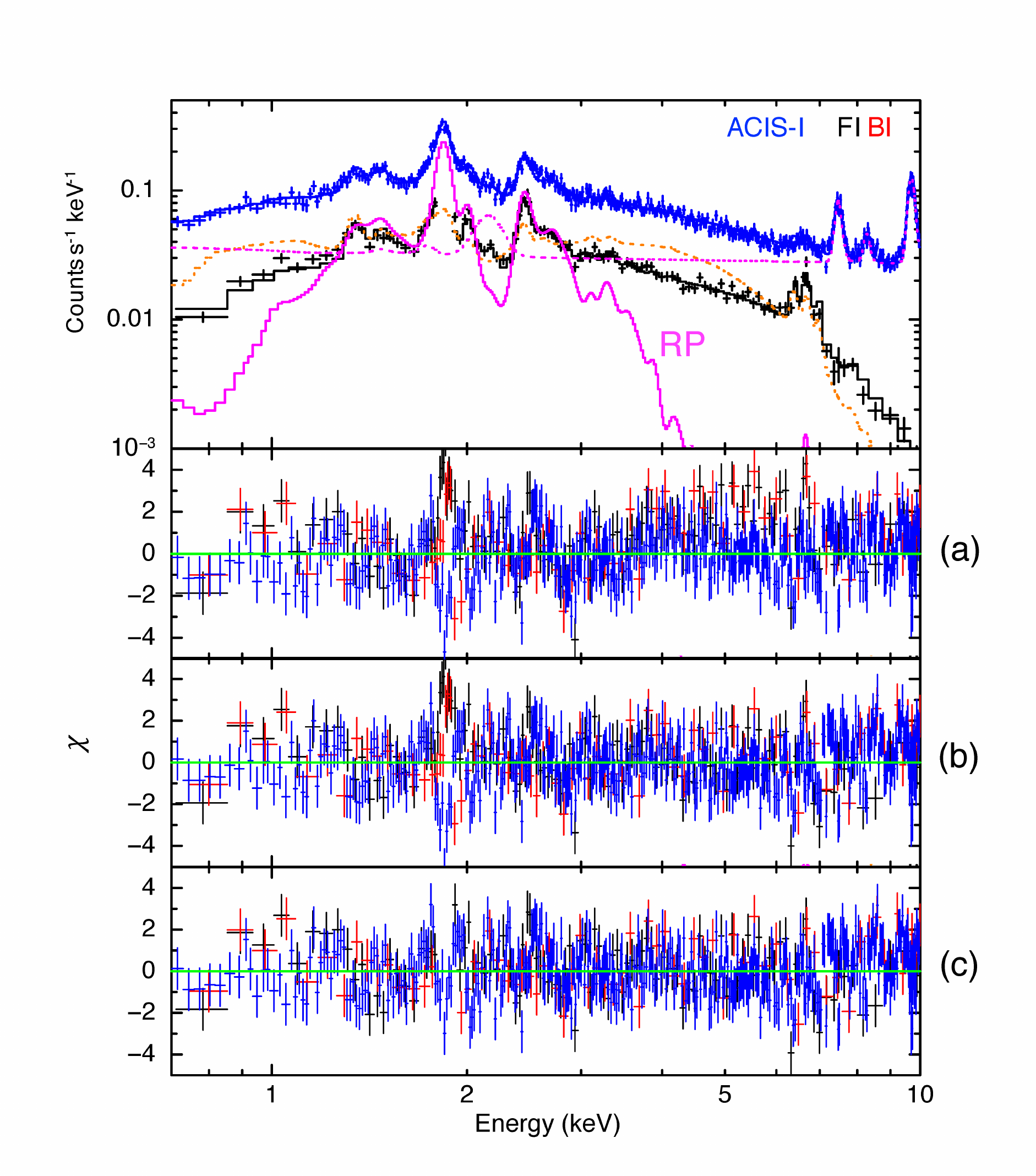}
\caption{The energy spectrum, best-fit model and residuals for the source-region spectra. The blue, black and red crosses and lines are used in the same way as in Figure \ref{spe_bgd_sc}. The orange and magenta dotted lines represent the Sky-BGD and NXB components for {\it Chandra} ACIS-I, whereas the magenta, solid line represents the RP component for {\it Chandra} ACIS-I. The residuals in the fittings with the models (a), (b) and (c) are shown in the panel (a), (b) and (c), respectively. \label{spe_src_sc}}
\end{figure}

First, we assumed that the Sky-BGD component had the same parameters as those of BGD region.
The NXB spectra for {\it Suzaku} was again reproduced with {\tt xisnxbgen} tool.
For {\it Chandra} data, we modeled the NXB spectrum as follows.
Since the intensity of the NXB varies depending on the position on the detector, the fluxes of the PL and Gaussian components in the NXB model were treated as free parameters, whereas the others were fixed to the best-fit values for the BGD data (see section \ref{bgd_spec}; hereafter, we refer to the model with this condition as ``model (a)'').
We fitted the data with model (a), and the residuals are shown in Figure \ref{spe_src_sc} (a). The reduced-$\chi^2 \,(d.o.f)$ was 2.16\,(571).
An excess of the data can be seen in the 4.0--7.0 keV energy range especially in the {\it Suzaku} data. Since the dominant emission in this energy range is the HP component (see Figure \ref{spe_bgd_sc}), we let ${\it EM}_{\rm HP}$ vary (hereafter, model (b)).
The residuals from the fit with model (b) are shown in Figure \ref{spe_src_sc} (b). The reduced-$\chi^2 \, (d.o.f)$ is 1.85 (570). The residuals in the 4.0--7.0 keV range were significantly reduced due to $\sim20\%$ higher normalization for the HP component.

The large residuals at $\sim$1.85 keV are probably due to the issues with the response matrices around the Si-K edge, since they show a large discrepancy between {\it Chandra} and {\it Suzaku} models, and it is unlikely that only the flux of the Si-line band differs significantly between {\it Chandra} and {\it Suzaku} data.
In fact, the quantum efficiency of the {\it Suzaku} XISs at $\sim$1.85 keV has a large uncertainty (\citealp{yamaguchi11}, {\it Suzaku} memo\footnote{ $\langle$ https://heasarc.gsfc.nasa.gov/docs/suzaku/analysis/sical\_update.html $\rangle$ }).
We therefore excluded the {\it Suzaku} data in 1.8--1.9 keV energy range (hereafter, model (c)). The result of the fitting with model (c) is shown in Figure \ref{spe_src_sc}. The spectra and best-fit models are shown in the top panel, and the residuals in the panel (c). This model fitted the data significantly better with a reduced-$\chi^2 \, (d.o.f)$ of 1.59\,(555). The best-fit parameters are presented in Table \ref{tab_src_sc}.

In order to check how the background parameter uncertainties affected the source parameters, we also conducted a simultaneous spectral fitting on the source and BGD region spectra using both {\it Chandra} and NXB-subtracted {\it Suzaku} data. In this analysis, we assigned the Sky-BGD and {\it Chandra} NXB components to the BGD region, and the Sky-BGD, {\it Chandra} NXB and source emission components to the source region.
The same Sky-BGD components except for the $EM_{\rm HP}$ were used in the source and BGD spectra. The $EM_{\rm HP}$ and {\it Chandra} NXB components were treated independently in the source and BGD region data. The free parameters were the same as those in the analysis in Sections \ref{bgd_spec} and \ref{spe_source}. The spectral model (c) was used to fit the source emission.
As a result, we obtained the plasma parameters, all of which were consistent with the values in Table \ref{tab_src_sc}. In particular, the $kT_{\rm e}$ and the lower limit of the $kT_{\rm init}$ were determined as 0.154 (0.138--0.171) keV and 5.7 keV.

Also, we conducted a background estimation using the spectrum extracted from the BGD2 region, which is outside the radio shell (see Figure \ref{img_suzaku}), in order to make sure that the background selection did not affect the source parameters significantly. Only the {\it Suzaku} data were available for the BGD2 region. We conducted spectral fitting on the BGD2 region using the {\it Suzaku} data and then source region using the {\it Chandra} and {\it Suzaku} data in the same way as described in Sections \ref{bgd_spec} and \ref{spe_source}. In this case, we also obtained the plasma parameters consistent with those in Table \ref{tab_src_sc}, including the $kT_{\rm e}$ and the lower limit of the $kT_{\rm init}$ of 0.141 (0.126--0.157) keV and 13.7 keV.

\begin{table*}[htb!]
\caption{ Best-fit parameters for the source-region spectra. }
\centering
\begin{threeparttable}
\begin{tabular}{l l c} \hline\hline
Component & Parameter\tnote{*} & Value \\ \hline

Source and Sky-BGD \\
RP& ${\it N}_{\rm H, RP}~{\rm (10^{22}~cm^{-2})}$ & 1.90 (1.76--2.06) \\
&${\it kT}_{\rm e}~{\rm (keV)}$ & 0.167 (0.153--0.192) \\
&${\it kT}_{\rm init}~{\rm (keV)}$ & $>$ 16 \\
&${\rm Mg~(solar)}$ & 0.8 (0.6--1.1) \\
&${\rm Si~(solar)}$ & 3.0 (2.6--3.6) \\
&${\rm S~(solar)}$ & 1.9 (1.6--2.4) \\
&$n_{\rm e} t$ (10$^{11}$ s~cm$^{-3}$)  & 4.1 (3.7--4.7) \\
&${\it EM}_{\rm RP}$ (10$^{-2}$) & 2.2 (1.7--2.8) \\
HP& ${\it EM}_{\rm HP}$ (10$^{-2}$) & 4.20 (4.12--4.29) \\
\\
NXB\\
Al-K$\alpha$&flux & $<$ 0.0003 \\
PL& Norm (10$^{-3}$)\tnote{\dag} & 7.90 (7.76--8.03) \\
Au-M$\alpha$& flux (10$^{-3}$) & 2.1 (1.8--2.4) \\
Ni-K$\alpha$ &flux (10$^{-3}$) & 3.01 (2.81--3.22) \\
Ni-K$\beta$& flux (10$^{-3}$) & 1.08 (0.92--1.25) \\
Au-L$\alpha$&flux (10$^{-3}$) & 7.1 (6.8--7.4) \\

& {\it Reduced-}$\chi^2 / {\it d.o.f}$ & 1.59/555 \\
\hline
\end{tabular}
\begin{tablenotes}\footnotesize
\item[*] The units of emission measures and Gaussian models are the same as those in Table \ref{tab_bgd_sc}.
\item[\dag] Normalization of the power-law model. The unit is the same as that in Table \ref{tab_bgd_sc}.

\end{tablenotes}
\end{threeparttable}
\label{tab_src_sc}
\end{table*}

\subsection{Spatially-resolved X-ray spectroscopy}

\begin{figure}[ht!]
\centering
\plotone{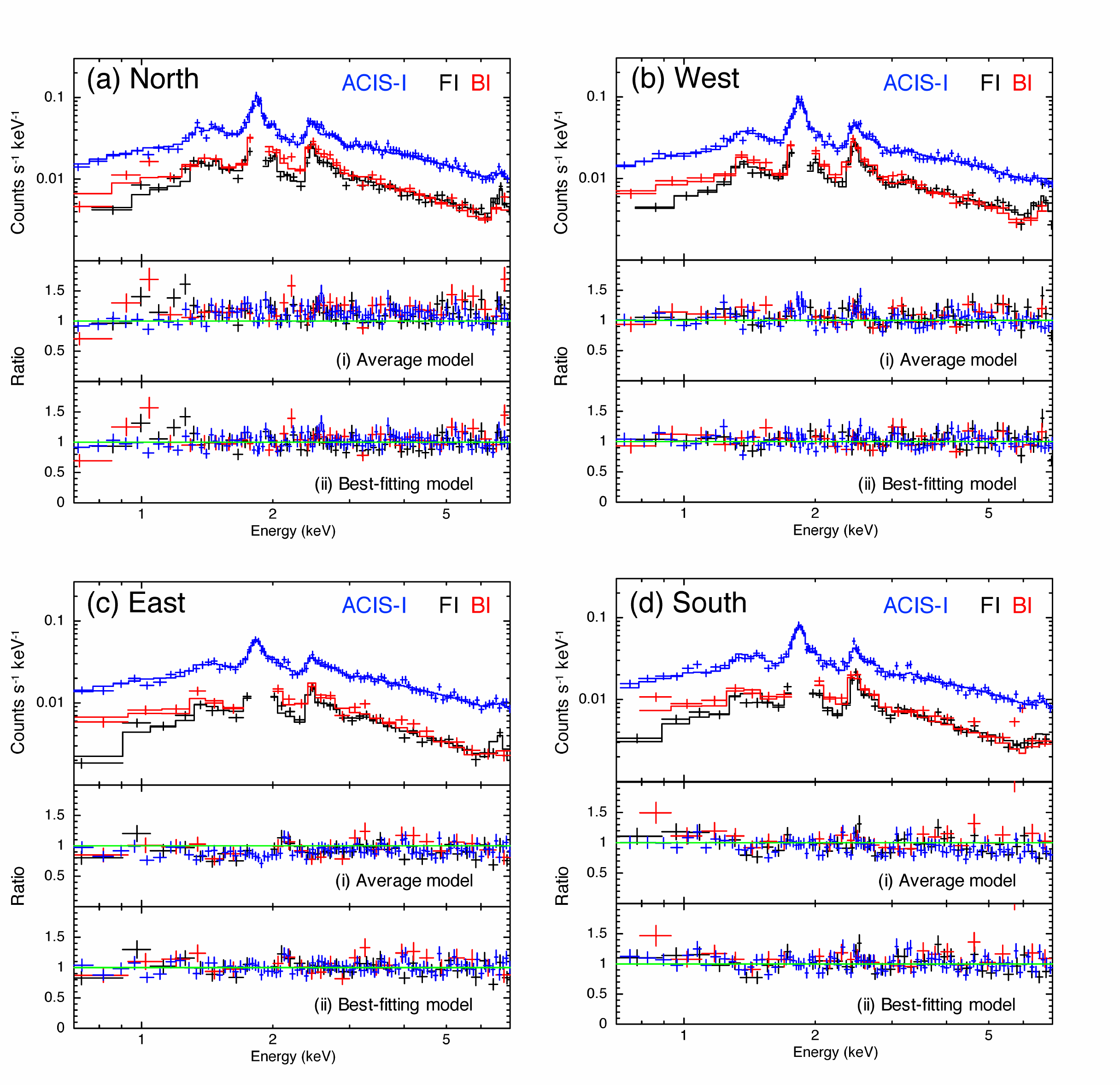}
\caption{The energy spectra, best-fit model spectra and ratios of the data and models for four regions. The blue, black and red crosses and lines are used in the same way as in Figure \ref{spe_bgd_sc}. In each of the panel (a)--(d), the data-to-model ratio using the average model (model (c) in section \ref{spe_source} multiplied by a quarter) and the best-fit model are shown. \label{spe_src4_sc}}
\end{figure}

We investigated spatial differences of the plasma parameters by extracting and fitting spectra from the four regions indicated in Figure \ref{img_chandra2}.
We fitted the spectra in 0.7--7.0 keV energy range, because the source-region data of {\it Suzaku} did not have sufficient statistics above 7 keV after we divided it into four regions.
First, we applied the model (c) scaled by the area ratios of each of the four regions and the whole region (hereafter, average model) to each spectrum. The residuals with the average model for each region are shown in the panel (i) in each of Figure \ref{spe_src4_sc} (a)--(d).
As indicated in Figure \ref{img_chandra2}, the North region is relatively bright and the East region has relatively low flux in the Si-line band.
Then, we fitted the data in each region with the model (c), but only $N_{\rm H}, kT_{\rm e}, kT_{\rm init}, n_{\rm e}t$, and $EM$ of the RP, $EM$ of the HP, and the fluxes of the Al-K$\alpha$, PL and Au-M$\alpha$ models of the NXB component were treated as free parameters (see Table \ref{tab_nwes}).

The data and best-fit spectra are shown in Figure \ref{spe_src4_sc}. The residuals are shown in the panel (ii) in each of Figure \ref{spe_src4_sc} (a)--(d).
The best-fit parameters are presented in Table \ref{tab_nwes}.
Only the emission measures of the RP components have significantly different values among four regions: the North and West regions have relatively high fluxes of the RPs. This trend is also consistent with Figure \ref{img_chandra2} (a) and (b).
Although the West region might have slightly lower $kT_{\rm e}$ than those of the other regions, the other parameters of the RP model did not show any significant variations among the four regions.

We also investigated the spectral variation with the SNR radius using the spectra extracted from the inner ellipse and outer oval ring shown in Figure \ref{img_chandra2} (d).
Conducting the same analysis as described in this section using the model (c) to fit the source emission, we did not find any significant parameter difference between the two regions.

\begin{table}[htb!]
\caption{ Best-fit parameters for each region. }
\centering
\begin{threeparttable}\tiny
\begin{tabular}{l l c c c c} \hline\hline
Component & Parameter\tnote{*} & North & West  & East & South \\ \hline
Source and Sky-BGD \\
RP &${\it N}_{\rm H, RP}~{\rm (10^{22}~cm^{-2})}$ & 1.81 (1.68--2.00) & 1.92 (1.75--2.15)  &  2.10 (1.85--2.43) & 1.69 (1.51--1.92) \\
&${\it kT}_{\rm e}~{\rm (keV)}$ & 0.18 (0.15--0.20) & 0.13 (0.11--0.15)  &  0.20 (0.16--0.24) & 0.15 (0.12--0.17) \\
&${\it kT}_{\rm init}~{\rm (keV)}$ & $>$ 12  &  $>$ 4.0 & $>$ 3.3  & $>$ 3.9 \\
&$n_{\rm e} t$ (10$^{11}$ s~cm$^{-3}$) & 4.1 (3.5--5.0) & 4.6 (3.8--5.5)  &  3.6 (2.7--4.7) & 3.8 (3.1--4.6) \\
&${\it EM}_{\rm RP}$ (10$^{-2}$) & 2.5 (1.9--3.4) & 3.7 (2.6--5.7)  &  1.6 (1.1--2.3) & 1.9 (1.4--2.7) \\
HP &${\it EM}_{\rm HP}$ (10$^{-2}$) & 4.91 (4.74--5.09) & 4.62 (4.46--4.70)  &  3.79 (3.59--3.99) & 4.08 (3.99--4.249) \\
\\
NXB\\
Al-K$\alpha$&flux & $<$ 0.0006  &  $<$ 0.0004  & $<$ 0.0005 & $<$ 0.0004 \\
PL &Norm (10$^{-3}$)\tnote{\dag} & 7.78 (7.43--8.13) & 7.44 (7.10--7.79)  &  7.20 (6.85--7.55) & 6.46 (6.12--6.78) \\
Au-M$\alpha$ &flux (10$^{-3}$) & 2.2 (1.7--2.8) & 1.8 (1.2--2.3)  &  2.2 (1.6--2.8) & 2.2 (1.7--2.8) \\
& {\it Reduced-}$\chi^2 / {\it d.o.f}$ & 1.32/206  &  1.13/181  & 0.98/149  &  1.78/161   \\
\hline
\end{tabular}
\begin{tablenotes}\footnotesize
\item[*] The units of emission measures and Gaussian models are the same as those in Table \ref{tab_bgd_sc}.
\item[\dag] Normalization of the power-law model. The unit is the same as that in Table \ref{tab_bgd_sc}.

\end{tablenotes}
\end{threeparttable}
\label{tab_nwes}
\end{table}

\subsection{Search for the compact remnant}

\begin{figure}[ht!]
\centering
\plotone{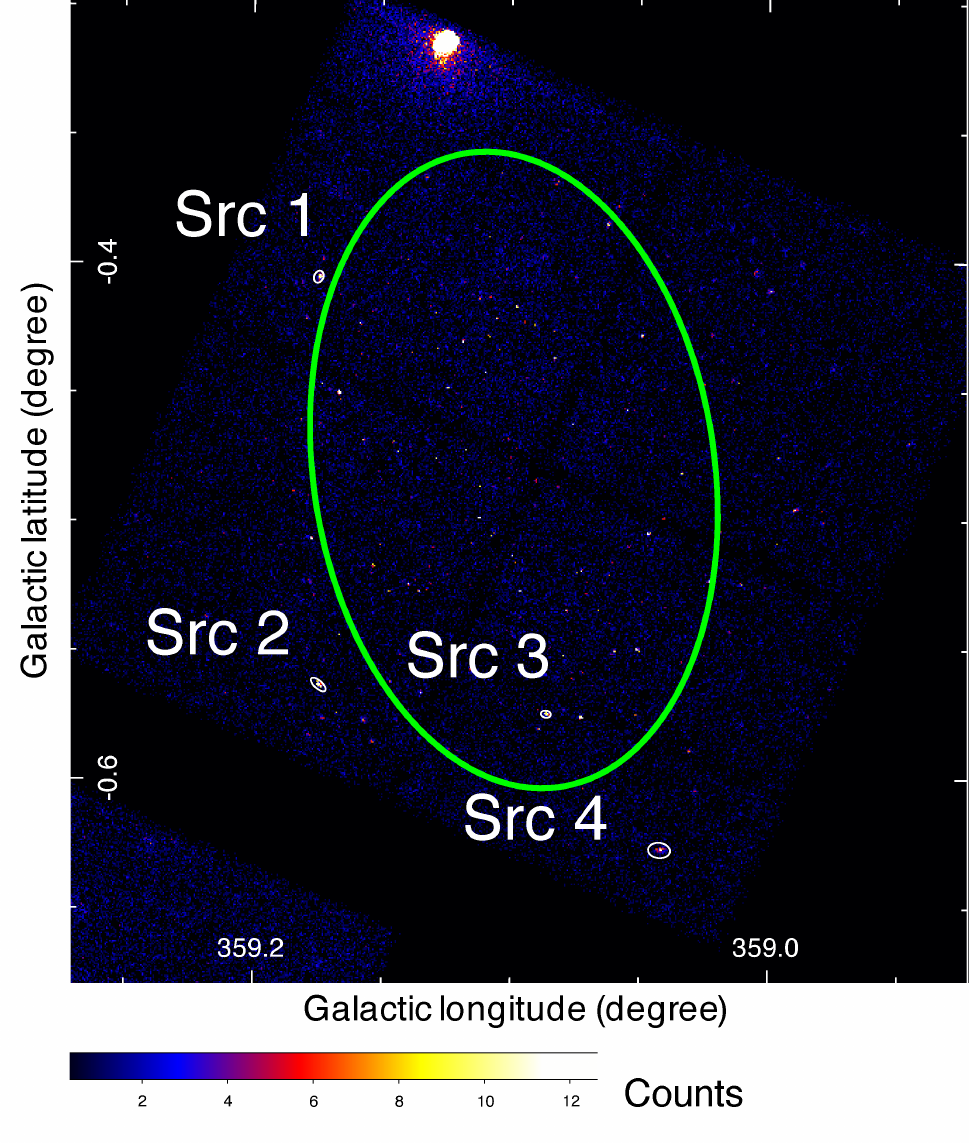}
\caption{The locations of the four brightest point-like sources around G359.1$-$0.5 are indicated with the white ellipses on the {\it Chandra} 0.7--5.0 keV X-ray image. The green large ellipse is the source region for G359.1$-$0.5. \label{img_points}}
\end{figure}

\begin{figure}[ht!]
\centering
\plotone{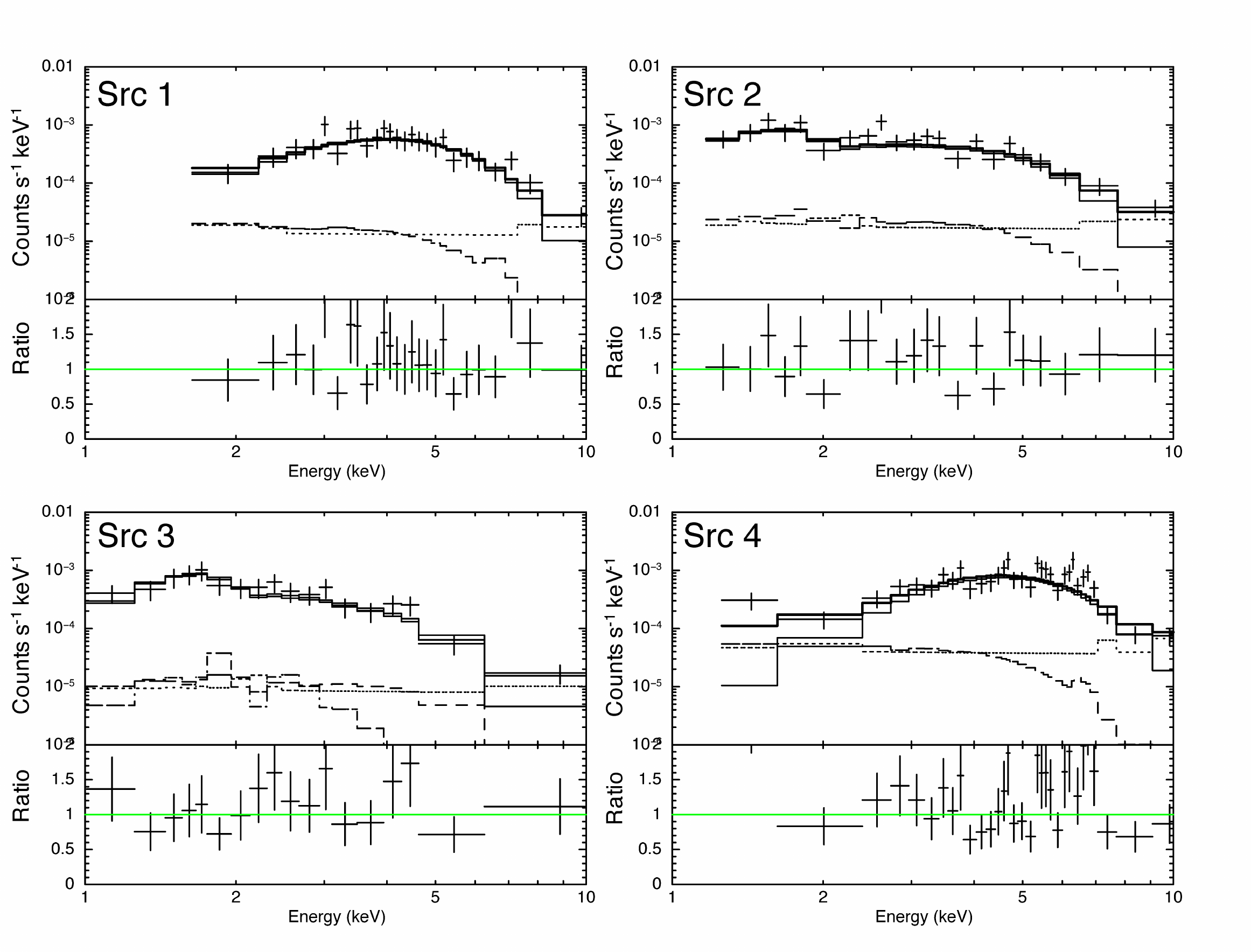}
\caption{The energy spectra of the four brightest point-like sources obtained with {\it Chandra} and the best-fit models. In each panel, the dotted, dashed and chain lines represent NXB, Sky-BGD and the RP components, respectively. The thin and thick solid line represent the absorbed power law and sum of the models, respectively. \label{spe_points}}
\end{figure}

In order to search for the compact remnant of G359.1$-$0.5, we first estimated total photon counts expected from the compact remnant of G359.1$-$0.5, assuming the distance of 4 kpc (see Section \ref{dist}) and the age of $>\,1\times10^4$ years (see Section \ref{prop}). Neutron stars with characteristic ages of (1--10)$\times 10^4$ years, a typical photon index of 2.0 \citep{kargal08}, and the same absorption column density as G359.1$-$0.5 ($2.0\times10^{22}$ cm$^{-2}$) typically have X-ray luminosities of $5 \times 10^{31}$--$10^{35}$ ergs s$^{-1}$ (\citealp{shibata16}; ATNF pulsar catalogue\footnote{$\langle$http://www.atnf.csiro.au/research/pulsar/psrcat/$\rangle$}), which correspond to the photon counts of $\sim100$--$10^{5}$ in our {\it Chandra} data.
Thus we extracted bright point-like sources around G359.1$-$0.5 with more than 100 counts in 0.7--5.0 keV energy band, from the ones detected with {\tt wavdetect} (except for The Great Annihilator at the northwestern corner of the ACIS-I; see Figure \ref{img_chandra} and \ref{img_points}).
The significance of the detection of four point-like sources were 12--18~$\sigma$.
In Figure \ref{img_points}, the $Chandra$ image (0.7--5.0 keV) and analysis regions for the spectroscopy of the selected sources are shown with the white ellipses.
The extracted spectra are shown in Figure \ref{spe_points}.
For the spectral analysis, we used the Sky-BGD model obtained in section \ref{bgd_spec}.
Also, we used the NXB model whose parameters were fixed to the best-fit values for the BGD region (see Table \ref{tab_bgd_sc}). 
As the source emission, we applied an absorbed power law model. The best-fit models and residuals are shown in Figure \ref{spe_points}. The best-fit parameters are presented in Table \ref{tab_points}.

\begin{table}[htb!]
\caption{Best-fit parameters for the point-like sources. \label{tab_points}}
\centering
\begin{threeparttable}\footnotesize
\begin{tabular}{l c c c c} \hline\hline
Parameter & Src 1 & Src 2 & Src 3 & Src 4 \\ \hline
${\it N}_{\rm H}~{\rm (10^{22}~cm^{-2})}$ & 5.02 (2.50--8.69)  &  0.69 (0.26--1.33)  & 1.58 (0.68--2.82)  &  5.94 (2.63--11.38) \\
${\Gamma}$ & 0.94 (0.10--1.94)  &  1.14 (0.65--1.70)  &  2.61 (1.77--3.63)  &  0.19 ($-0.87$--0.60) \\
Norm$_{\rm PL}$ (10$^{-6}$)\tnote{*} & 8.49 (1.85--46.2)  &  5.56 (2.78--12.2)  & 23.5 (6.94--95.4)  &  2.58 (0.75--11.2) \\
{\it Reduced-}$\chi^2 / {\it d.o.f}$ & 0.73/23 & 1.06/20  & 0.85/16 & 1.58/31 \\
\hline
\end{tabular}
\begin{tablenotes}\footnotesize
\item[*] Normalization of the power-law model. The unit is the same as that in Table \ref{tab_bgd_sc}.

\end{tablenotes}
\end{threeparttable}
\end{table}

\subsection{Distribution of molecular clouds \label{co}}

\begin{figure}[!htbp]
\epsscale{0.95}\plotone{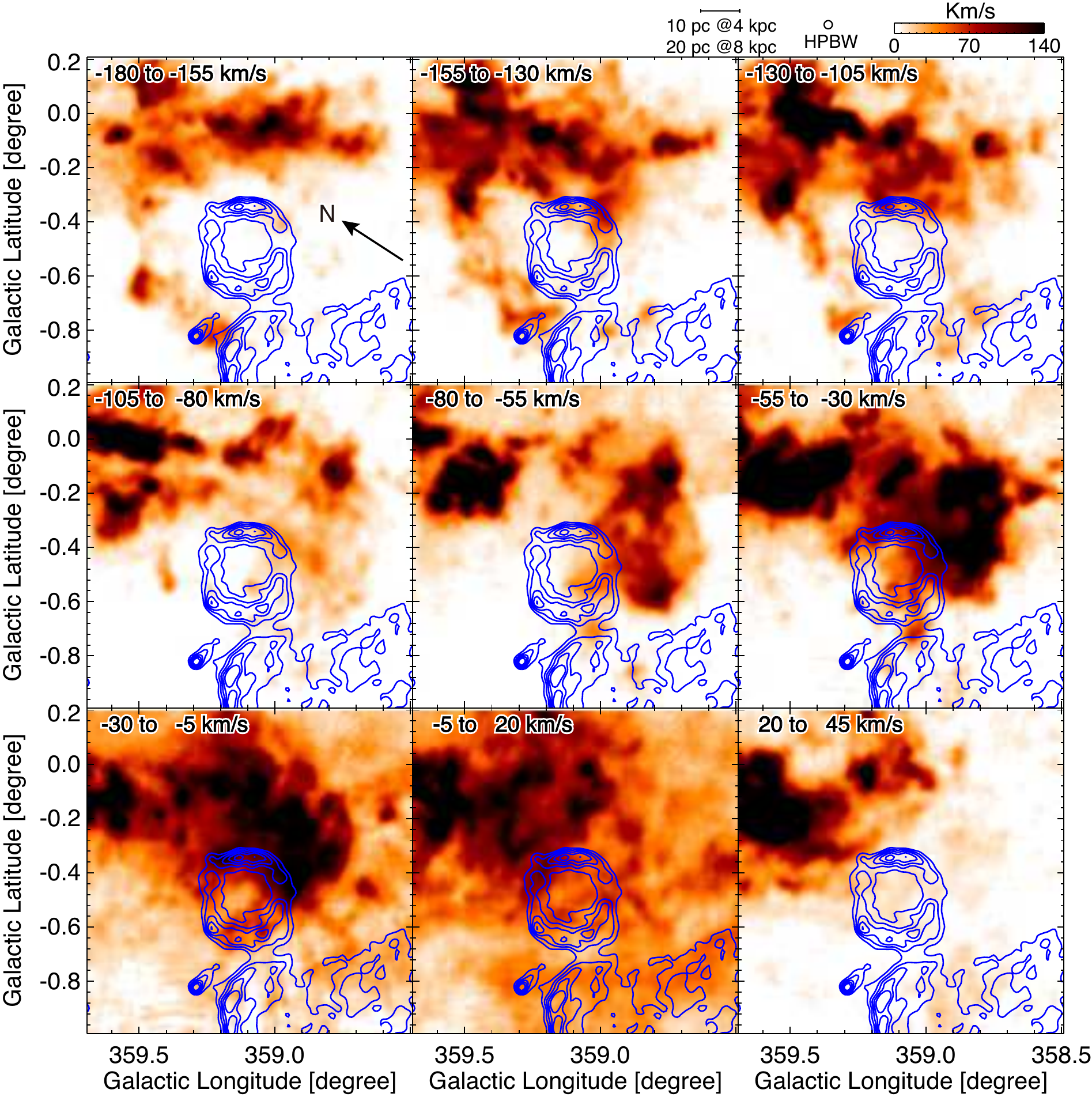} 
\caption{Velocity channel distribution of the $^{12}$CO($J$=2--1) emission integrated every 25 km~s$^{-1}$ overlaid with the radio continuum emission at 60 cm as blue contours.
\label{fig:lbch}}
\end{figure}

 Figure \ref{fig:lbch} shows the velocity channel distribution of $^{12}$CO($J$=2--1) toward G359.1$-$0.5 with the blue contours of the radio continuum emission obtained by \citet{lar00}. In $V_{LSR}$ = $-180$ to $-55$ km~s$^{-1}$, two molecular clouds are seen in the vicinity of the SNR; one is a filamentary cloud located in the east to the north of the SNR extending from the Galactic plane and the other is a diffuse cloud located in the south. \citet{uchida92a} proposed that these clouds are associated with the SNR. However, our data with the finer angular resolution revealed that the spatial distribution between the clouds and the SNR did not necessarily exhibit a good correspondence. In particular, the panel at $V_{LSR}$ = $-155$ to $-130$ km~s$^{-1}$ shows an obvious offset between the filamentary cloud and the SNR.

\begin{figure}[!htbp]
\epsscale{0.95}\plotone{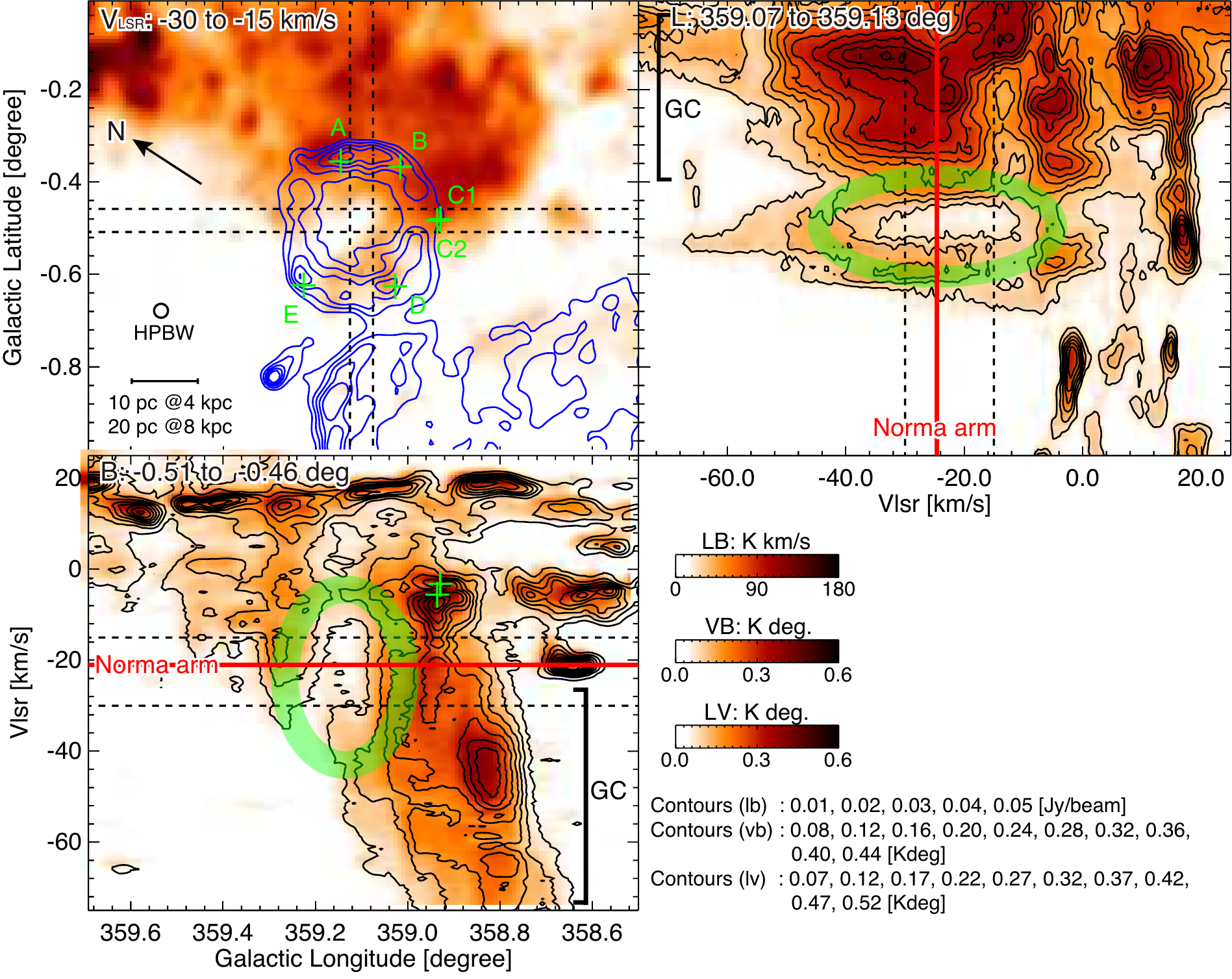} 
\caption{($top-left$) Integrated intensity distribution of the $^{12}$CO($J$=2--1) emission of the molecular cloud associated with G359.1$-$0.5. The blue contours indicate the radio continuum shell of G359.1$-$0.5 at 60 cm. Light green crosses indicate the positions of OH masers obtained by \citep{yusef95}. The dashed lines indicate the integration range of $l$ and $b$ for the $v-b$ and $l-v$ diagrams. ($top-right$) Velocity-Latitude diagram of $^{12}$CO($J$=2--1). The red line indicates the mean velocity of clouds in the Norma arm toward the integrating Longitude. Dashed lines indicate the velocity integration range of the $l-b$ map. ($bottom-left$) Longitude-Velocity diagram of $^{12}$CO($J$=2--1). The red line indicates the mean velocity of clouds in the Norma arm toward the integrating Latitude. Dashed lines indicate the velocity integration range of the $l-b$ map. The green transparent rings in top-right and bottom-left panels highlight a shell-like structure in the CO data.
\label{fig:lbv}}
\end{figure}

We carefully reviewed the CO data and finally found another candidate of the associated cloud at $V_{LSR}$ $\sim$$-20$ km~s$^{-1}$. Figure \ref{fig:lbv} shows Longitude-Latitude, Velocity-Latitude, and Longitude-Velocity diagrams toward the candidate cloud. Light green crosses indicate the positions of OH masers obtained by \citet{yusef95}. We found a shell-like structure represented as green transparent rings in the position-velocity diagrams and found that the OH masers are located at the outer boundary of the cloud.
This clearly indicates the interaction between the expanding CO structure and the radio shell, as they show the consistent sizes with each other.

\section{Discussion}

We carefully investigated the ionization state of the plasma of G359.1$-$0.5, and estimated the recombination timescale $n_{\rm e} t$ and the initial temperature $kT_{\rm init}$ for the first time. We did not find any significant spatial variation of the plasma parameters.
We also analyzed the CO data in the vicinity of the SNR, and discovered a new and most plausible candidate for the associated clouds.

\subsection{Estimation of the distance to G359.1$-$0.5}\label{dist}

First, we discuss the distance of the associated CO cloud which was found in section \ref{co}.
In Figure \ref{fig:lbv}, the red lines indicate the intensity-weighted mean velocities of clouds in the Norma arm in the integrated $l$ and $b$ ranges. The Velocity-Latitude and the Longitude-Velocity diagrams display that the velocity center of the shell-like structure corresponds to the velocity of the Norma arm.
As a result, we suggest that G359.1$-$0.5 is located in the Norma arm with the distance of $\sim$4 kpc \citep{rei16}, though we still can not completely exclude the possibility that the cloud is located in the GC region with the central velocity corresponding to the Norma arm by chance.
By using the conversion factor from the intensity of $^{12}$CO($J$=1--0) to the column density of hydrogen ($N_\mathrm{H_2}$) derived by \citet[][1.0 $\times 10^{22}$ cm$^{-2}$ K km~s$^{-1}$]{oka17} and our $^{12}$CO($J$=1--0) data, we derive $N_\mathrm{H_2}$, total molecular mass, and kinetic energy of the associated cloud (see Table \ref{tab:mc}).

Second, from our X-ray analysis, the absorption column density $N_{\rm H}$ toward the SNR was estimated as $\sim$1.8$\times10^{22}$ cm$^{-2}$, which is consistent with the value derived in the previous work \citep{ohnishi}, and also comparable to that of the nearby pulsar, Mouse \citep{mori05}. The Mouse is believed to be located at the distance of 3--5 kpc \citep{camilo02}.
These results are consistent with previous work (\citealp{yusef95}; \citealp{lazendic02}).

According to these results, we conclude that the most reasonable distance to G359.1$-$0.5 is $\sim$4 kpc.
Hereafter, we assume the distance to the SNR to be 4 kpc.

\begin{table}[h!]
\centering
\caption{Physical parameters of the molecular cloud} \label{tab:mc}
\begin{threeparttable}
\begin{tabular}{lcccc}
\tablewidth{0pt}
\hline
\hline
Distance (kpc) & $N_\mathrm{H_2}$ (cm$^{-2}$) & Mass (M$_\odot$) & Expansion velocity (km~s$^{-1}$)\tnote{*} & E$_{kin}$ (erg)\\
\hline
\decimals
4.0 & 5.6 $\times$ 10$^{22}$ & $\sim$3 $\times$ 10$^{5}$ & 20 & $\sim$8 $\times$ 10$^{50}$    \\
8.3 & 3.9 $\times$ 10$^{22}$ & $\sim$6 $\times$ 10$^{5}$ & 20 & $\sim$3 $\times$ 10$^{51}$    \\
\hline
\end{tabular}
\begin{tablenotes}\footnotesize
\item[*] Velocity separation between the red-shifted part and the blue-shifted part of the shell structure.

\end{tablenotes}
\end{threeparttable}
\end{table}

\subsection{Physical parameters of the G359.1$-$0.5 plasma \label{prop}}

The average electron density of the plasma is calculated as $\sim$0.57$f^{-0.5}$ cm$^{-3}$ from the emission measure and the volume, assuming an ellipsoid of 8.7, 5.2, 5.2 pc and a volume filling factor $f$. Using this value, we estimate the plasma mass and total thermal energy as $\sim$6$f^{0.5}\,M_\odot$ and $\sim$1.4 $\times$ 10$^{48}f^{0.5}$ ergs, respectively. The recombination age can also be estimated as $\sim$2.4 $\times$ $10^{4}f^{0.5}$ years, which is rather small among known recombining SNRs \citep{suzuki18}.
Assuming the Sedov model \citep{sedov59} with an explosion kinetic energy of 10$^{51}$ ergs, current shock radius of 14~pc and uniform ISM density of 0.57$f^{-0.5}$/4 $\,=\,0.14 f^{-0.5}$~cm$^{-3}$, we also estimate the lower limit of the SNR age as 1$\times$10$^{4} f^{0.1}$ years, which is consistent with the estimated recombination age. We note that, since the SNR has the centrally-filled X-ray structure and contains the RP, it should have had dense surrounding CSM and/or clouds, which slowed down its evolution. Thus our estimation here should be the lower limit of its age.

\begin{figure}[ht!]
\centering
\plotone{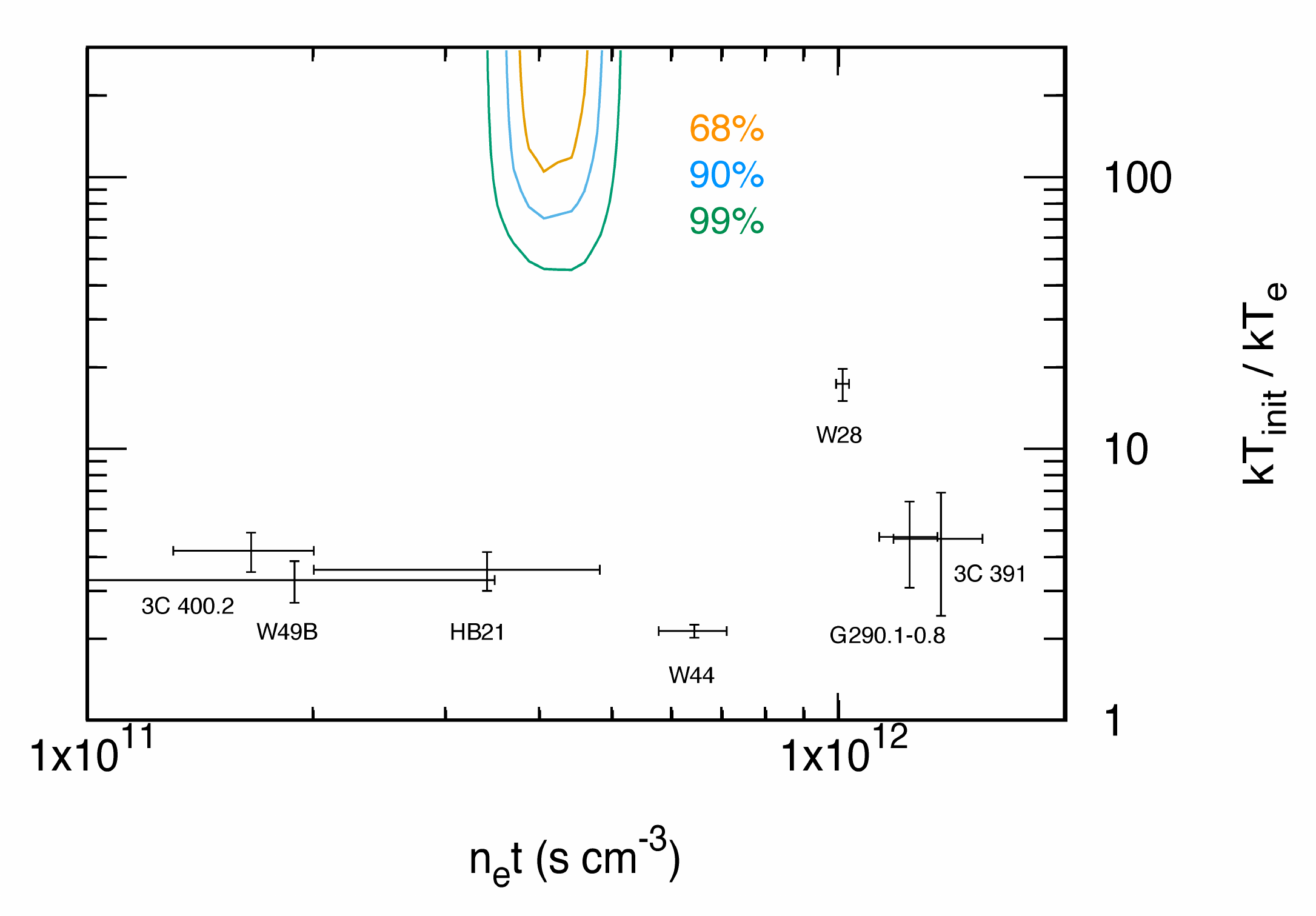}
\caption{The ${\chi^2}$ contours for the parameter space of ${\it n_{\rm e}t}$ and ${\it kT_{\rm init}/kT_{\rm e}}$. The orange, cyan, and green lines represent 68\%, 90\%, and 99\% confidence levels, respectively. The black crosses represent values of the other SNRs (3C 400.2: \citealp{ergin17}; W49B: \citealp{yamaguchi18}; HB21: \citealp{suzuki18}; W44: \citealp{uchida12}; W28: \citealp{okon18}; G290.1-0.8: \citealp{kamitsukasa15}; 3C 391: \citealp{sato14}). \label{contours}}
\end{figure}

Despite the large physical size of the SNR, the plasma parameters show surprisingly little spatial variation. We estimate the sound crossing time of the current SNR plasma to be
\begin{equation}
8.7 \times 10^{4}~{\rm years} \,({\rm length}/15~{\rm pc}) (kT_{\rm e}/0.17~{\rm keV})^{-0.5},
\end{equation}
which is longer than or comparable to the SNR age ($>$ 1$\times 10^{4}$ years).
This implies that substantial ionization and subsequent rapid cooling took place in the early evolutionary stage, when the remnant size was still small.
The ratio of $kT_{\rm init}$ and $kT_{\rm e}$ ($>\,\sim$70) is significantly higher than that of the other evolved recombining SNRs (see Figure \ref{contours} and text below), in most of which thermal conduction to surrounding cloudlets is suggested as an origin of rapid cooling. We note that this high ratio is not due to the background selection bias as we confirmed in Section \ref{spe_source}. If we take into account the background parameter uncertainties or take the other background region for spectral fitting, we obtain this ratio of $> \sim 37$, which is still prominent among known recombining SNRs (Figure \ref{contours}). Although this ratio is extraordinarily high, the value of $> \sim 50$ is in fact explained assuming a simple rarefaction scenario in a certain condition as described in \cite{itomasai}.
Furthermore, with the remarkably circular shape of the radio shell and detection of OH masers from several points, the SNR seems to have begun interacting with MCs rather recently.
Although this requires dense CSM around the SNR and low-density region surrounding the CSM, both of them can be expected if the progenitor is red supergiant (e.g. \citealp{dwark05}). Additionally, with the low-density cavity region, the uniform plasma properties we figured out are also naturally explained.
Therefore, as for the formation process of RPs, we conclude that the rarefaction scenario is more plausible.

We compare the plasma state of G359.1$-$0.5 to the other recombining SNRs. Note that we exclude the SNRs which were analyzed with fixed $kT_{\rm init}$ (e.g. IC~443: \citealp{matsumura18}).
Figure \ref{contours} shows recombination timescales ${\it n_{\rm e}t}$ and ratios of initial temperatures $kT_{\rm init}$ and electron temperatures $kT_{\rm e}$ of the SNRs. The contours show the confidence regions for G359.1$-$0.5. G359.1$-$0.5 has a remarkably large ratio of the $kT_{\rm init}$ and $kT_{\rm e}$, indicating that it is still in a highly recombination-dominant state, although its ${\it n_{\rm e}t}$ is a typical value among the SNRs. The contours are elongated along the vertical axis because of the large uncertainty of the $kT_{\rm init}$ (see Table \ref{tab_src_sc}).



\subsection{Association of the interacting CO clouds with GeV/TeV emission}

\begin{figure}[ht!]
\centering
\plotone{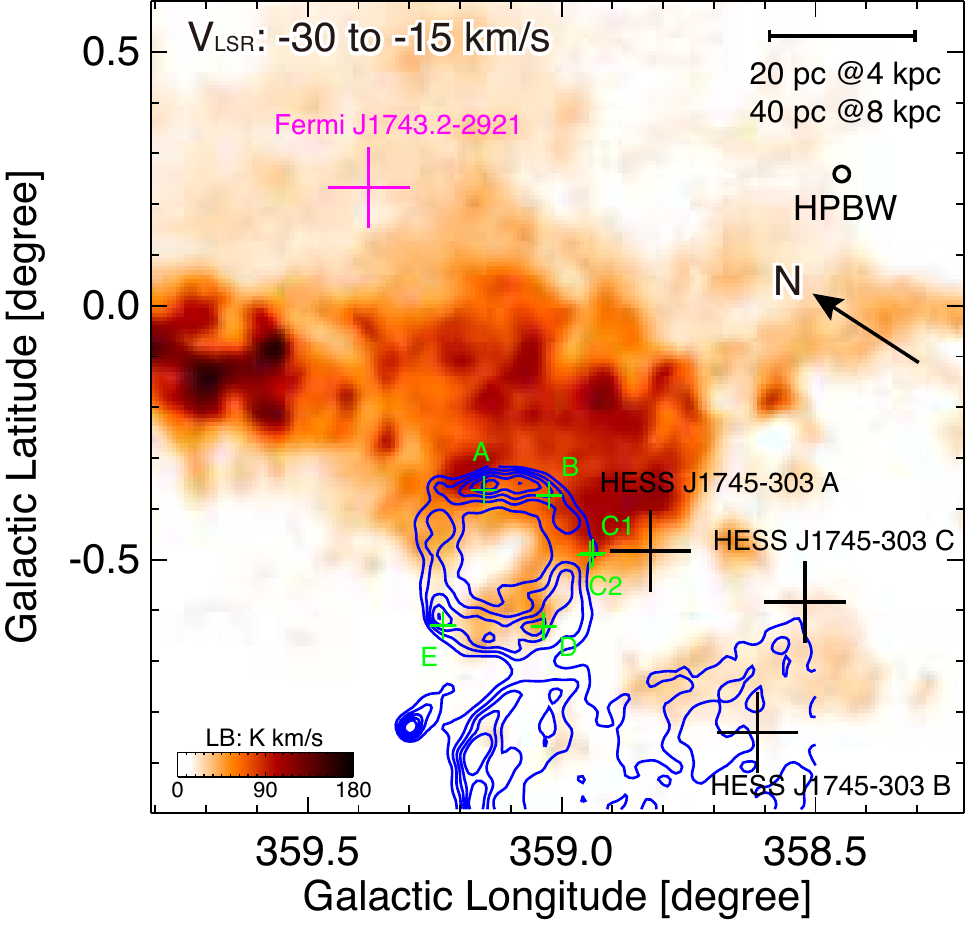}
\caption{The distribution of the $^{12}$CO($J$=2--1) emission of the molecular cloud associated with G359.1$-$0.5 and the locations of the GeV/TeV sources. The image and blue contours are same as those in the top-left panel of Figure \ref{fig:lbv}. The black and magenta crosses represent the peak positions of the sources detected with H.E.S.S and {\it Fermi}. \label{img_co_gamma}}
\end{figure}

Here, we investigate the association of the CO cloud we detected in section \ref{co} with GeV/TeV emission.
Figure \ref{img_co_gamma} shows comparison between the distribution of the associated CO clouds and locations of GeV/TeV sources.
The peak positions of the H.E.S.S sources are shown with the three black crosses (HESS J1745$-$303 A, B, and C) \citep{aharo08}.
The peak positions of the {\it Fermi} sources are indicated with the black cross (HESS J1745$-$303 A) and magenta cross (Fermi J1743.2$-$2921) \citep{hui16}, as well.
No clear spatial coincidences between these sources and the CO cloud found in this work are seen.

Regarding the H.E.S.S sources, \cite{hayakawa12} suggested that CO and HI clouds located in the GC with the velocity range of -100 to -40 km/s, which is significantly different from those of the associated cloud of G359.1-0.5 found in this work ($\sim-20$ km s$^{-1}$), were responsible for the TeV emission.
Also, \cite{bamba09} reported the association of neutral iron line possibly from cold material such as MCs and TeV gamma-rays, but Figure \ref{img_co_gamma} shows no association of clouds and TeV emission. It indicates that the detected neutral iron line originates from clouds at a different distance. It is consistent with the fact that the neutral iron emission is possibly due to the irradiation of MCs at the GC to past X-ray flares of the GC.

Fermi J1743.2$-$2921 has elongated spatial feature extending from the SNR \citep{hui16} and we cannot completely rule out the possibility of association with the SNR.
Some mixed morphology SNRs with a separation of $\sim$10 pc between the gamma-ray emission and the radio shell have been explained with the ``leaked cosmic ray model'', in which leaked cosmic rays from the broken shell emit GeV/TeV gamma-rays (e.g. W28: \citealp{cui18}). However, the peak position of the Fermi J1743.2$-$2921 is $\sim$40 pc away from the SNR in projection and thus may be difficult to be explained.

Consequently, we find no clear spatial/velocity coincidence between GeV/TeV emission and the CO found in this work, although Fermi J1743.2$-$2921 cannot be ruled out as unrelated.
For more precise study, gamma-ray observations with better angular resolution (e.g. Cherenkov Telescope Array: CTA; \citealp{hermann08}), as well as millimeter observations to search for the clouds being ionized by cosmic rays, will be needed.

\subsection{Origins of the bright point-like sources}

We detected four point-like sources as the candidates of the compact remnant of G359.1$-$0.5 (see Figure \ref{spe_points}, Table \ref{tab_points}).
Comparing the absorption column densities to that of G359.1$-$0.5, Src 3 is most likely to be located in the vicinity of the SNR.
Considering the best-fit power-law indices, Src 1--3 have typical values of pulsars \citep{kargal08}, while Src 4 exhibits too hard a spectrum.
Comparing also to the properties of the pulsar wind nebula associated with the SNR W44 (photon index of $\sim$2.2, X-ray luminosity of $\sim 6\times10^{32}$ ergs s$^{-1}$; \citealp{petre02}), Src 1--3 have similar values of photon indices and slightly lower X-ray luminosities.
According to these results, in our dataset, we conclude that Src 3 is the most plausible candidate of the compact remnant of the SNR.

Assuming Src 3 to be the compact remnant and the SNR age of $>$ 1$\times$10$^4$ years (see section \ref{prop}), its velocity in the plane of the sky is estimated as $<$ 500 km s$^{-1}$.
We note that the ``Mouse'' (G359.23$-$0.82) is also one of the compact remnant candidates, and it has a similar velocity of $\sim$300 km s$^{-1}$ assuming the distance of 4 kpc (\citealp{camilo02}; \citealp{mori05}).

\section{Conclusion}

We conducted a spatially-resolved X-ray analysis of the SNR G359.1$-$0.5 using the {\it Chandra} and {\it Suzaku} archival data.
We properly estimated the ionization state of the plasma for the first time by carefully modeling the background emission and the {\it Chandra} NXB spectra. We found that the deviation of $kT_{\rm e}$ ($\sim$0.17 keV) from the initial temperature $kT_{\rm init}$ ($>$16 keV) was remarkably large among the recombining SNRs, although its recombination timescale ${\it n_{\rm e} t}$ is typical for these sources (4.2 $\times$ 10$^{11}$ s cm$^{-3}$).

In order to confirm the distance to the SNR, we also conducted an analysis of CO clouds in the vicinity of the SNR. Using high spatial-resolution data obtained with NANTEN and NANTEN2, we discovered a new and most plausible candidate of the associated clouds. The cloud has a line-of-sight velocity of $\sim$$-20$ km s$^{-1}$, indicating that the distance to the SNR is $\sim$4 kpc. This result is consistent with the value estimated from the absorption column density obtained from our X-ray analysis. Therefore, we conclude that the distance to G359.1$-$0.5 is $\sim$4 kpc.
As for a formation process of the RP, we favor the rarefaction scenario because of the remarkably high $kT_{\rm init}$ and large sound crossing time.

A comparison between spatial distribution of the associated CO cloud and that of GeV/TeV emission showed no clear spatial coincidence with each other, although the GeV emission cannot be totally ruled out as unrelated.
As a candidate of the compact remnant of the SNR, a plausible point-like source which is located at $\sim$6' away from the SNR center, was discovered.

\acknowledgements{

We thank the anonymous referee for a plenty of useful advice, especially regarding the justifications of the extraordinarily high initial temperature of the source.
HS is deeply grateful to H. Matsumura for a discussion on thermal conduction processes in plasmas, and S. Tsujimoto for useful information on H\,{\footnotesize I} study and OH maser physics.
HS is supported by JSPS Research Fellowship for Young Scientist (No. 19J11069).
This work is supported by the Chandra GO Program grant GO2-13098X, the Grant-in-Aid for Scientific Research on Innovative Areas “Toward new frontiers: Encounter and synergy of state-of-the-art astronomical detectors and exotic quantum beams” (18H05459; AB), JSPS/MEXT KAKENHI grant No. 19K03908 (AB), and Shiseido Female Researcher Science Grant (AB).
PPP acknowledges support under NASA contract NAS8-03060 with the Chandra X-ray Center.
}

\software{XSPEC (v12.9.1; \citealp{arnaud96}), HEAsoft (v6.20; \citealp{heasarc14}), xisnxbgen \citep{tawa08}, CIAO (v4.10; \citealp{fruscione06})}




\end{document}